\documentclass[AMS,STIX2COL,SYSTEMFONTS]{WileyNJD-v2}
\usepackage{moreverb}

\usepackage{graphicx}
\usepackage{lineno}
\usepackage[T1]{fontenc} 

\usepackage{amsmath,amsfonts,amsthm,bm,bbm}
\usepackage{tabularx}
\usepackage{tabularray}[=v2021]
\usepackage{multirow}
\UseTblrLibrary{booktabs}

\usepackage{color, colortbl}
\definecolor{Gray}{gray}{0.9}

\usepackage{algorithm}
\usepackage{afterpage}
\usepackage{color,soul}
\usepackage{realboxes}
\usepackage{multirow}
\usepackage{hhline}
\usepackage{tikz}
\usepackage{float}
\usetikzlibrary{bayesnet}
\usepackage{comment}
\usepackage{natbib}

\articletype{Article Type}%

\received{<day> <Month>, <year>}
\revised{<day> <Month>, <year>}
\accepted{<day> <Month>, <year>}

\begin{document}

\title{Bayesian Posterior Interval Calibration to Improve the Interpretability of Observational Studies}
%\runtitle{Bayesian Posterior Interval Calibration for Observational Studies}

\author[1,2]{Jami J. Mulgrave*}

\author[1,4]{David Madigan}

\author[1,2,3]{George Hripcsak}

\authormark{MULGRAVE \textsc{et al}}

\address[1]{ \orgname{Observational Health Data Sciences and Informatics (OHDSI)}, \orgaddress{\state{New York}, \country{USA}}}

\address[2]{\orgdiv{Department of Biomedical Informatics}, \orgname{Columbia University}, \orgaddress{\state{New York}, \country{USA}}}

\address[3]{\orgdiv{Medical Informatics Services}, \orgname{New York-Presbyterian Hospital}, \orgaddress{\state{New York}, \country{USA}}}

\address[4]{\orgdiv{Khoury College of Computer Sciences}, \orgname{Northeastern University}, \orgaddress{\state{New York}, \country{USA}}}

\corres{*Jami Mulgrave, \email{jnj2102@gmail.com}}

%\presentaddress{Present address}

\abstract[Abstract]{Observational healthcare data offer the potential to estimate causal effects of medical products on a large scale.  However, the confidence intervals and p-values produced by observational studies only account for random error and fail to account for systematic error. As a consequence, operating characteristics such as confidence interval coverage and Type I error rates often deviate sharply from their nominal values and render interpretation impossible. While there is longstanding awareness of systematic error in observational studies, analytic approaches to empirically account for systematic error are relatively new. Several authors have proposed approaches using negative controls (also known as ``falsification hypotheses'') and positive controls. The basic idea is to adjust confidence intervals and p-values in light of the bias (if any) detected in the analyses of the negative and positive control. In this work, we propose a Bayesian statistical procedure for posterior interval calibration that uses negative and positive controls.  We show that the posterior interval calibration procedure restores nominal characteristics, such as 95\% coverage of the true effect size by the 95\% posterior interval.}


\maketitle

%\footnotetext{\textbf{Abbreviations:} ANA, anti-nuclear antibodies; APC, antigen-presenting cells; IRF, interferon regulatory factor}

%% ** Mainmatter **

\section{Introduction}\label{fig:sec1}
Observational healthcare data offer the potential to estimate causal effects of healthcare interventions at scale.  However, because of concerns about bias and the appropriate interpretation of the statistical artifacts produced by observational studies, healthcare researchers, practitioners, and regulators have struggled to incorporate observational studies into routine healthcare decision making. Observational studies must consider two sources of error, random and systematic. Random error derives from sampling variability whereas systematic error derives from confounding, measurement error, model mis-specification, and related concerns. Standard statistical methods generally focus exquisitely narrowly on random error and assume there is no systematic error. Ironically,  random error generally converges to zero as sample size become larger, precisely the circumstance we now enjoy with large-scale electronic health records.  Systematic error, by contrast, persists independently from sample size and thus increases in relative importance. Negative controls, i.e., exposure-outcome pairs where one believes no causal effect exists, have been proposed as a tool to better explain systematic error  \citep{lipsitch_negative_2010, tchetgen_tchetgen_control_2014, dusetzina_control_2015, arnold_brief_2016, desai_utilization_2017}.  Executing a study on negative controls and determining whether the results indeed show no effect, that is using them as ``falsification hypotheses'', can help detect bias inherent to the study design or data.  In order to account for these biases,  \cite{schuemie_empirical_2018} and \cite{schuemie_improving_2018} go one step further and incorporate the effect of error observed for negative controls into the estimates from observational studies, in effect calibrating the confidence intervals and p-values.  Their methods can also account for positive controls, exposure-outcome pairs where the (non-null) causal effect is known at least approximately. 

In this paper, we derive and analyze a Bayesian hierarchical analog to frequentist bias correction or "calibration" approaches described in \cite{schuemie_empirical_2018} and \cite{schuemie_improving_2018}. The core idea is to harness large sets of negative controls along with a key exchangeability assumption to maintain nominal statistical properties such as Type I error rate. We present a large-scale evaluation in two different clinical contexts, depression and hypertension.  There are connections between our work and prior papers that have developed Bayesian hierarchical Gaussian models for meta-analysis that incorporate bias correction \citep{welton_models_2009, dias_estimation_2010, verde_biascorrected_2021, raices_cruz_robust_2022}, although these prior approaches do not focus on negative controls.  Our Bayesian approach clarifies the assumptions underlying calibration and facilitates future extensions such as calibrated meta-analysis and calibrated model averaging. 

\section{Methods} 

\subsection{Basics}
We define a
cohort $c, c = 1,\ldots, C$ as a group of subjects that satisfy one or
more criteria for a duration of time. For example, a cohort could
comprise individuals newly diagnosed with hypertension, with one
year's observation prior to cohort entry, prescribed a beta blocker at
cohort entry, and followed thereafter for two years. A subject
can belong to multiple cohorts at the same time, and belong to the
same cohort multiple times. For example, a subject could drop in and
out of a hypertension cohort according to whether they are taking or
not taking a particular drug. We refer to each period of time a
subject is in a cohort as an ``entry.''  We denote by $N_c$ the number
of entries in cohort $c$ and by $d_{ci}$ the duration (in days) for entry
$i$ in cohort $c$.  Finally, $N_c (t)$ denotes the number of entries in cohort
$c$ that span day $t$.

We use cohorts to study associations between interventions and
``outcomes.'' An outcome (e.g., stroke) occurs at a discrete moment in
time and may or may not have a duration. We denote by $y_{ci}$ the
number of outcome events observed for entry $i$ in cohort $c$ and by
$y_{ci} (t)$ the number of outcome events observed for entry $i$ in
cohort $c$ on day $t$.

An ``exposure cohort'' is a cohort where all entries are exposed to a
particular treatment $x, x = 1,\ldots, X$. As such, $y_{ci} (x=j)$
denotes the number of outcome events for subject $i$ in exposure
cohort $c$ associated with treatment $x = j$.  We can also consider a
counterfactual cohort, identical in every way, except each subject is
unexposed to treatment $j$ at all times while in the cohort.
Here. $y_{ci} (x=\lnot j)$ denotes the number of outcome events for patient
$i$ in this counterfactual cohort.

We define the causal effect of $x = j$ on $Y$, within some exposure cohort
$c$ defined by exposure to $x = j$, as the incidence rate ratio:
\[
\mu_{cx} = \frac{\sum_i y_{ci}(x=j)/\sum_i d_{ci}}{\sum_i
  y_{ci}(x=\lnot j)/\sum_i d_{ci}}.
\]

Alternatively, we can also formulate the effect as the hazard ratio: 
\[
h_{cx}(t) = \lim_{\Delta t \to 0} \frac{  \sum_i y_{ci}
    (x=j,[t,t+\Delta t])/N_c(t) }{ \sum_i y_{ci} (x=\lnot j,[t,t+\Delta t])/N_c(t)}.
\]
%Since $c$ is defined by factually being exposed to
%$x = 1, y_{ci} (x=1)$ will be completely observed, and our counterfactual
%$y_{ci} (x=0)$ will be completely unobserved. 
Note that these quantities estimate the average treatment
effect in the treated (ATT) for a cohort. 

Finally, $y_{ci} (x=j,t)$ denotes the number of outcome events on day
$t$ for subject $i$ in exposure cohort $c$ associated with treatment
$x = j$ and $y_{ci} (x=\lnot j,t)$ denotes the corresponding quantity in the
counterfactual unexposed cohort.

\subsection{Negative and Positive Controls}

A ``negative control'' comprises of a drug and an outcome, where we believe that the drug does not cause the outcome.  In this work, we use these negative controls in pairs in which we consider both a target treatment and a comparator treatment, in order to use the data that we are analyzing found in the comparative effectiveness study  \citep{schuemie_empirical_2018}.  These data will be described in greater detail in Subsection \ref{datadescr}.  Thus, for this work, a negative control comprises of a
target treatment, a comparator treatment, an outcome, and nesting cohort where
neither the target treatment nor the comparator treatment are believed
to cause the corresponding outcome. Therefore, the true effect sizes
for the direct causal effect of the target treatment on the outcome,
the direct causal effect of the comparator on the outcome, and the
comparative effect of the target treatment versus the comparator
treatment on the outcome are all 1. For example, consider the
first row of Table~\ref{Gold}, and let $y$ denote the outcome
``acute pancreatitis'' and $x=j$ denote the treatment brinzolamide.  Because we assume we have a negative control, the relative risk must be 1, and we therefore have
\[ 
y_{ci} (x=j,t)=y_{ci} (x=\lnot j,t), i=1,\ldots,N_c, t = 1,\ldots,d_{ci}.
\]
As a consequence, both $\mu_{cx}=1$ and $h_{cx}(t)=1.$

\begin{table*}[htp]
\centering
\begin{tabular}{ccccc} 
 \hline
 \hline
Target	& Comparator	& Nesting cohort	&Outcome	&True effect size\\ \hline
\rowcolor{Gray} Brinzolamide	&Levobunolol	&Glaucoma	&Acute pancreatitis	&1.0\\
Cevimeline	&Pilocarpine	&Sjogren's syndrome	&Acute pancreatitis	&1.0\\
\rowcolor{Gray} Diclofenac	  &Celecoxib	&Arthralgia	&Acute stress disorder	&1.0\\
Diclofenac 	&Celecoxib	&Arthralgia	&Ingrowing nail	&1.0\\
\hline
\end{tabular}
\caption{Example of negative controls.}
\label{Gold}
\end{table*}

We generate synthetic ``positive controls'' from negative controls 
by adding simulated additional outcomes in the
target treatment cohort until a desired incidence rate ratio is
achieved. For example, assume that, during treatment with diclofenac,
$m$ occurrences of ingrowing nail were observed. None of these were
caused by diclofenac since this is one of our negative controls. If we
were to add an additional $m$ simulated occurrences during treatment with
diclofenac, we would have doubled the observed effect size. Since this was a
negative control, and since  only the treatment cohort received
new ingrowing nails and not the counterfactual cohort, the observed
relative risk which was one becomes two. 

More specifically, let $\theta$ denote the target effect
size. Currently we use $\theta=1.5, \theta=2$ and $\theta=4$ to
generate 3 positive controls from every negative control. We increase
outcome count $y_{ci}(x=j)$ in the target treatment ($j$) cohort to
$y_{ci}^* (x=j)$ to approximate the desired $\theta$. To avoid issues
due to low sample size, we generate positive controls only when
$\sum_i y_{ci} \geq 25.$

The steps in the ``injection'' process are as follows:
\begin{enumerate}
\item Within the target treatment cohort $c$, we fit an L1-regularized
  Poisson regression model where the outcome
  $y_{ci}(x=j)$ represents the subject-level dependent variable and
  $Z_{ci}$ represents the independent variables. The independent
  variables include demographics, as well as all recorded diagnoses,
  drug exposures, measurements, and medical procedures all measured
  prior to cohort entry (``baseline covariates''). We use 10-fold
  cross-validation to select the regularization hyperparameter. Let
  $\hat\lambda_{ci}=E(y_{ci} |Z_{ci})$ denote the predicted Poisson
  event rate for entry $i$ in treatment cohort $c$.
\item For every entry in the target treatment cohort, sample $n$ from
  a Poisson distribution with parameter $(\theta-1)\hat\lambda_{ci}$
  and set $y_{ci}^* (x=j)= y_{ci}(x=j)+n.$
\item Repeat step 2 until $\lvert \frac{\sum_iy_{ci}^*(x=j)/\sum_i
    d_{ci}}{\sum_i y_{ci}(x=j)/\sum_i d_{ci}}-\theta \rvert \leq \epsilon,$ where
  $\epsilon$ is currently set to 0.01.
\end{enumerate}

Assuming the synthetic
outcomes have the same measurement error (same positive predictive
value and sensitivity) as the observed outcomes, this process creates
data that mimic a true marginal effect size of $\theta$. Because we sample
new outcomes from a large-scale predictive model, we also mimic the
conditional effect (conditional on $Z$). We note that the altered data can
capture effects due to measured confounding but not unmeasured
confounding.

\subsection{Frequentist Calibration}

In prior work, we described a method for empirically calibrating
p-values \citep{schuemie_interpreting_2014}.  Briefly, when evaluating a particular analytical
method, the calibration procedure applies the method not only to the
target-comparator-outcome of interest but also to
all the negative controls. This generates draws from an ``empirical'' null
distribution. By contrast with the theoretical null distribution
(typically a Gaussian centered on 1), the empirical null distribution
does not assume that the estimated effect size provides an unbiased
estimate of the true effect. Instead the location and dispersion 
of the empirical null distribution reflects both random error and 
systematic error. ``Calibrated'' or ``empirical'' p-values use the
empirical null distribution in place of the theoretical null
distribution when computing p-values. 

More formally, let $\hat\theta_i$ denote the estimated log effect
estimate (relative risk, odds or incidence rate ratio) from the $i$th
negative control, and let $\hat\tau_i$ denote the corresponding
estimated standard error, $i=1,\ldots,n$.  Let $\theta_i$ denote the
true log effect size (assumed 0 for negative controls), and let
$\beta_i$ denote the true (but unknown) bias associated with pair $i$,
that is, the difference between the log of the true effect size and
the log of the estimate that the study would have returned for control
$i$ had it been infinitely large. As in the standard p-value
computation, we assume that $\hat\theta_i$ is normally distributed
with mean $\theta_i + \beta_i$ and variance $\hat\tau_i^2$.  Note that
in traditional p-value calculations, $\beta_i$ is assumed to be equal
to zero for all $i$. Instead we assume the $\beta_i$'s arise from a
normal distribution with mean $\mu$ and variance $\sigma^2$. This
represents the null (bias) distribution. We estimate $\mu$ and
$\sigma^2$ via maximum likelihood. In summary, we assume the
following:
\[
\beta_i \sim N(\mu,\sigma^2) \: {\rm and} \: \hat\theta_i \sim N(\theta_i+\beta_i,\hat\tau_i^2),
\]
where $N(a,b)$ denotes a normal distribution with mean $a$ and variance $b$, and we estimate $\mu$ and $\sigma^2$ by maximizing:
\[
\prod_{i=1}^n \int p(\hat\theta_i \mid \beta_i, \theta_i,
\hat\tau_i^2)p(\beta_i \mid \mu,\sigma^2)d\beta_i
\]
yielding maximum likelihood estimates $\hat\mu$ and $\hat\sigma^2$.
We compute a calibrated p-value that uses the empirical null
distribution. Let  
$\hat\theta_{0}$ denote the log of the effect estimate for the outcome
of interest, and let $\hat\tau_{0}$  
denote the corresponding (observed) estimated standard error. Assuming  
$\beta_{n+1}$ arises from the same null distribution, we have that:
\[
\hat\theta_{0} \sim N(\hat\mu , \hat\sigma^2 + \hat\tau_{0}^2),
\]
and the p-value calculation follows naturally.

A later paper \citep{schuemie_empirical_2018} used positive controls to extend
the concept of calibrated p-values to calibrated confidence intervals.
These intervals reflect actual accuracy on negative and
positive controls and, like calibrated p-values, capture both random
and systematic error. More specifically, they assumed that $\beta_i$, the bias associated
with control $i$, again comes from a Gaussian distribution, but this 
time using a mean and standard deviation that are linearly related to  
$\theta_i$, the true effect size:
\[
\beta_i \sim N(\mu(\theta_i),\sigma^2(\theta_i))
\]
where:
\[
\mu(\theta_i)=a + b \theta_i  \; {\rm and} \; 
\sigma^2(\theta_i)=c+d \times \mid\theta_i\mid.
\]
We estimate $a,b,c$ and $d$ by maximizing the marginalized likelihood
in which we integrate out the unobserved $\beta_i$:
\[
\prod_{i=1}^n \int p(\hat\theta_i\mid \beta_i,\theta_i,\hat\tau^2_i)
p(\beta_i \mid a,b,c,d,\theta_i)d\beta_i
\] 
yielding $(\hat{a},\hat{b},\hat{c},\hat{d})$. 
We compute a calibrated CI that uses the systematic error model. Let $\hat\theta_{0}$
again denote the log of the effect estimate for the outcome of
interest, and let $\hat\tau_{n+1}$ denote the corresponding (observed)
estimated standard error. Then:
\[
\hat\theta_{0} \sim N(\theta_{0} +\hat{a} + \hat{b} \times
\theta_{0},\hat{c}+\hat{d} \times \mid \theta_{0} \mid + \hat\tau^2_{0}),
\]
and the calibrated confidence interval follows.

Our prior work has also shown that, unlike
standard confidence intervals, calibrated confidence intervals
maintain coverage at or close to the desired level, e.g., 95\%.
Typically, but not necessarily, the calibrated confidence interval is
wider than the nominal confidence interval, reflecting the problems
unaccounted for in the standard procedure (such as unmeasured
confounding, selection bias and measurement error) but accounted for
in the calibration. Since the uncertainty about the bias cannot be negative, the coverage of the calibrated interval is higher. 

Our approach to calibration is agnostic to the method used to generate the estimates $\hat{\theta}_i, i=0,\ldots,m$. However, it does assume that the {\it same} method is used to generate $\hat{\theta}_0$ (that is, the estimate of the effect size of interest), as is used to generate the estimates, $\hat{\theta}_i, i=1,\ldots,m$, that correspond to the negative and positive controls. Note that $\hat{\theta}_i, i=1,\ldots,m$, are ``estimating'' a known quantity since for the negative and positive controls the corresponding parameters $\theta_i, i=1,\ldots,m$, are assumed to be known.

\subsection{Bayesian Calibration}
In this work, we use a hierarchical Bayesian approach to perform the calibration.  We consider two different models of the bias, a model that is not dependent on the true effect sizes, and a model that is linearly dependent on the true effect sizes. %The model assumes that the true bias follows a Gaussian distribution around the true effect size.  

Again, let $\theta_0$ denote the log of the true effect size (relative risk, odds ratio, or hazard ratio) of interest.  Let $\theta_i$ denote the log of the true effect size for the available negative and positive controls, where $i = 1, \ldots ,m$.  We use $\hat{\theta}_i$ to denote the log of the estimated effect sizes, where $i = 0, \ldots, m$, and $\hat{\tau_i}$ to denote the corresponding estimated standard  error. In what follows, we will again assume that the $\beta_i$'s are normally distributed. In practice, as shown below, this yields good empirical results but future refinements could consider more flexible models.

\subsubsection{Constant Model of the True Bias}
\label{ConstantModel}
Let $\hat{\beta}_i = \hat{\theta}_i - \theta_i $ denote the ``estimated bias,'' where $i = 0, \ldots,m$.  

Then we propose the following hierarchical model:
\begin{align}
\beta_i &\sim N(\mu, \sigma^2), \; i = 0, \ldots, m \\
\hat{\beta}_i &\sim N(\beta_i, \hat\tau_i^2), \; i = 0, \ldots, m 
\end{align}

Note in the above model, it is possible to integrate out the $\beta_i$'s since:
\begin{align}
\hat{\beta}_i \sim N(\mu, \sigma^2 + \hat\tau_i^2) 
\end{align}

\begin{figure}[h!]
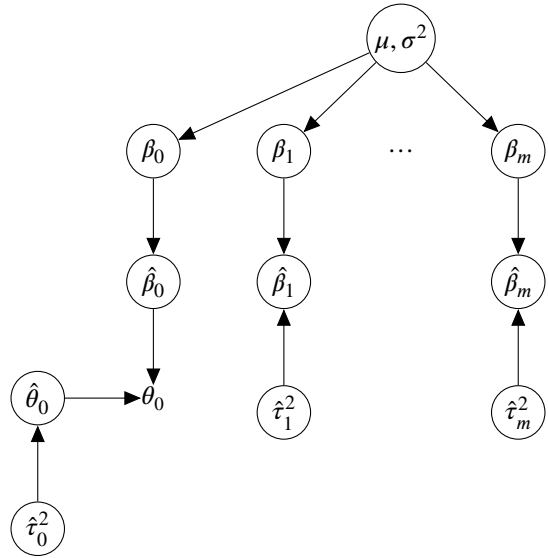

\caption{Graphical Model for Constant Model of the True Bias.}
\label{fig:SimpleModel}
  \centering
  \tikz{ %

    \node[latent] (beta0) {$\beta_0$} ; %
    \node[latent, right=of beta0] (beta1) {$\beta_1$} ; %
    \node[const, right=of beta1] (dots){$\ldots$} ; %
    \node[latent, right=of dots] (betam) {$\beta_m$} ; %

    \node[latent, above=of dots] (mu) {$\mu,\sigma^2$} ; %

    \edge {mu}{beta0} ; %
    \edge {mu}{beta1} ; %
    \edge {mu}{betam} ; %

    \node[latent, below=of beta0] (betahat0){$\hat\beta_0$} ; %
    \node[latent, right=of betahat0] (betahat1) {$\hat\beta_1$} ; %
    \node[latent, below=of betam] (betahatm) {$\hat\beta_m$} ; %

    \edge {beta0}{betahat0} ; %
    \edge {beta1}{betahat1} ; %
    \edge {betam}{betahatm} ; %

    \node[const, below=of betahat0] (theta0){$\theta_0$} ; %
    \node[latent, left=of theta0] (thetahat0){$\hat\theta_0$} ; %

    \edge {betahat0}{theta0} ; %
    \edge {thetahat0}{theta0}
    
    \node[latent, below=of thetahat0] (tausquared0){$\hat\tau^2_0$} ; %

        \node[latent, below=of betahat1] (tausquared1){$\hat\tau^2_1$} ; %
\node[latent, below=of betahatm] (tausquaredm){$\hat\tau^2_m$} ; %

\edge {tausquared0}{thetahat0};
    \edge {tausquared1}{betahat1} ; %
    \edge {tausquaredm}{betahatm}

  }
\end{figure}

We give $\mu$ and $\sigma^2$ relatively noninformative priors, $\mu \sim N(0, 5)$ and $1/\sigma^2 \sim U(0, 100)$, where $U$ stands for the uniform distribution.  See Figure \ref{fig:SimpleModel} for a graphical model representation.

\subsubsection{Linear Model of the True Bias}
\begin{figure}[h!]
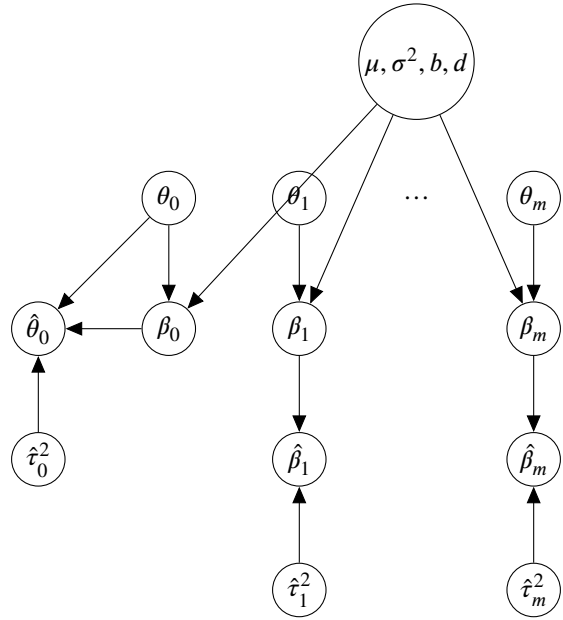

\caption{Graphical Model for Linear Model of the True Bias.}
\label{fig:LinearModel}
  \centering
  \tikz{ %

    \node[latent] (theta0) {$\theta_0$} ; %
    \node[latent, right=of theta0] (theta1) {$\theta_1$} ; %
    \node[const, right=of theta1] (dots){$\ldots$} ; %
    \node[latent, right=of dots] (thetam) {$\theta_m$} ; %

    %\node[latent] (beta0) {$\beta_0$} ; %
    %\node[latent, right=of beta0] (beta1) {$\beta_1$} ; %
    %\node[const, right=of beta1] (dots){$\ldots$} ; %
    %\node[latent, right=of dots] (betam) {$\beta_m$} ; %

    \node[latent, above=of dots] (mu) {$\mu,\sigma^2, b, d$} ; %

   % \edge {mu}{theta0} ; %
%    \edge {mu}{theta1} ; %
 %   \edge {mu}{thetam} ; %

    \node[latent, below=of theta0] (beta0){$\beta_0$} ; %
    \node[latent, below=of theta1] (beta1) {$\beta_1$} ; %
    \node[latent, below=of thetam] (betam) {$\beta_m$} ; %

    \edge {mu}{beta0} ; %
    \edge {theta0}{beta0} ; %

    \edge {mu}{beta1} ; %
    \edge {theta1}{beta1} ; %
    
    \edge {mu}{betam} ; %
    \edge {thetam}{betam} ; %

    \node[latent, left=of beta0] (thetahat0){$\hat\theta_0$} ; %
    \node[latent, below=of beta1] (betahat1) {$\hat\beta_1$} ; %
    \node[latent, below=of betam] (betahatm) {$\hat\beta_m$} ; %

    \edge {beta0}{thetahat0} ; %
    \edge {beta1}{betahat1} ; %
    \edge {betam}{betahatm} ; %

    \edge {theta0}{thetahat0} ; %

    \node[latent, below=of thetahat0] (tausquared0){$\hat\tau^2_0$} ; %

        \node[latent, below=of betahat1] (tausquared1){$\hat\tau^2_1$} ; %
\node[latent, below=of betahatm] (tausquaredm){$\hat\tau^2_m$} ; %

\edge {tausquared0}{thetahat0};
    \edge {tausquared1}{betahat1} ; %
    \edge {tausquaredm}{betahatm}

  }
\end{figure}

We compare the results of the constant model described in \ref{ConstantModel} of the true bias to the linear model described in \cite{schuemie_empirical_2018}.  Following \cite{schuemie_empirical_2018}, we reconstruct the linear model of true bias of using the modified hierarchical model:

\begin{align}
    \beta_i \sim N(\mu + c\theta_i, \sigma^2 + d\theta_i), \; i = 0, \ldots, m
\end{align}

where $\mu {\sim} N(0, 5)$, $1/{\sigma}^2  {\sim} U(0, 100)$, $c {\sim} N(0,5),$ and $d {\sim} U(0, 100)$.  See Figure \ref{fig:LinearModel} for a graphical model representation.

\subsubsection{Experimental Design and Computation}

We consider two approaches to evaluating Bayesian calibration.  For the first approach, we train the Bayesian calibration method on 80\% of the treatment-comparator-outcome (TCO) triples using both positive and negative controls, and test the effect size of interest using the remaining TCOs.  Since the positive controls are simulated from the negative controls, we take care to test the positive and negative controls that were not used in the same TCO combinations from the training data set.  We evaluate both models of bias, the simple model and the linear model, using this training data set.

For the second approach of evaluating Bayesian calibration, and to address concerns about the fidelity of the positive controls with respect to unmeasured confounding and the actual true effect sizes, we train the Bayesian calibration method on 80\% of the negative control combinations, and test the positive and negative controls that were not used in the same TCO combinations from the training data set.  Due to the nature of the linear model, we can only evaluate the constant bias model using this training data set.

We use Markov chain Monte Carlo, specifically a Gibbs sampler, to estimate the log of the effect sizes for the Bayesian calibration procedure.  We use R with the JAGS program to define the prior distribution and run the
MCMC.  We specify three parallel chains, 1000 burn-in and adaptive iterations, thinning of one, and 1000 additional samples to take. Initial values for each chain are specified to have a mean of zero and a variance of one.  The trace, histogram, empirical cumulative distribution function, and autocorrelation plots reflect adequate convergence.  We sample $\beta$ and find the posterior intervals by finding the 95th quantiles of the posterior samples of $\theta$, denoted in the code as `thetaTest.'
\\

For the constant model, we use the following BUGS syntax, where nu = $1/(\sigma^2)$. Note that JAGS uses precision in the normal distribution:
\begin{lstlisting}[basicstyle=\footnotesize]
model{
for(i in 1:Ntrain){
  betaHat[i] ~ dnorm(mu, 1/(1/nu + 1/tauHat[i]))  
}
for(i in 1:Ntest){
  thetaTest[i] <- thetaTestHat[i] - betaTestHat[i]
  betaTestHat[i] ~ dnorm(mu,1/(1/nu + 1/tauTestHat[i]))  
}
mu ~ dnorm(0,0.2)
nu ~ dunif(0,100)
}
\end{lstlisting}

For the linear model, we use the following BUGS syntax:
\begin{lstlisting}[basicstyle=\footnotesize]
model{
for(i in 1:Ntrain){
  betaHat[i] ~ dnorm(mu+(c*theta[I]),  
               1/(1/nu + (d*theta[i]) + 1/tauHat[i]))   
}
for(i in 1:Ntest){
  thetaTest[i] ~ dnorm(0,0.2)
  betaTest[i] ~ dnorm(mu+(c*thetaTest[I]),
                1/(1/nu+(d*thetaTest[i])))  
  thetaTestHat[i] ~ dnorm(thetaTest[i]+betaTest[I],
                    tauTestHat[i])  
}
mu ~ dnorm(0,0.2)
c ~ dnorm(0,0.2)
d ~ dunif(0,100)
nu ~ dunif(0,100)
}
\end{lstlisting}

\subsubsection{Description of the Data}\label{datadescr}
We consider two data sets for the analysis, one with treatments for depression and another with treatments for hypertension.  

For the depression data set, we used the results described in \cite{schuemie_improving_2018}.   
\cite{schuemie_improving_2018} reported a large-scale comparative effectiveness study of 17 treatments for depression with respect to 22 outcomes in four large-scale observational databases. Using a new-user propensity-adjusted cohort design, they reported 5,984 estimated effect sizes and corresponding nominal and calibrated 95\% confidence intervals. Negative controls were identified for the 17 treatments by selecting outcomes that are well studied, but for which no evidence in the literature suggests a relationship with any of the 17 treatments. A total of 52 negative control outcomes were selected \citep{schuemie_improving_2018}.  Since real positive controls for observational research are difficult to obtain in practice, \cite{schuemie_improving_2018} generated synthetic positive controls by modifying a negative control through injection of additional, simulated occurrences of the outcome.  The injected extra outcome occurrences were drawn from a high-dimensional patient-level predictive model. For each negative control that has a relative risk of 1, three positive controls were generated, with relative risks of 1.5, 2, and 4. The four databases included in the study for the depression data set were 1) Truven MarketScan Multi-state Medicaid (MDCD), 2) OptumInsight's de-identified Clinformatics Datamart (Optum), 3) Truven MarketScan Medicare Supplemental Beneficiaries (MDCR), and 4) Truven MarketScan Commercial Claims and Encounters (CCAE).  These databases and the procedure for selecting negative and positive controls are described in detail in \cite{schuemie_improving_2018}.

Similar procedures were employed for the generation of the hypertension data set.  The hypertension treatments consists of six administrative claims databases and three electronic health record databases.  The databases are: 1) IBM MarketScan
Multi-state Medicaid (MDCD), 2) Optum ClinFormatics (Optum), 3) IBM
MarketScan Medicare Supplemental Beneficiaries  (MDCR),  4) IBM
MarketScan Commercial Claims and Encounters (CCAE), 5) Japan Medical Data Center (JMDC), 6) IMS/Iqvia Disease Analyzer Germany (IMSG), 7) Optum Pan-Therapeutic (Panther), 8) Korea National Health Insurance Service / National
Sample Cohort (NHIS NSC), and 9) Columbia University Medical Center (CUMC). These databases and the generation of the negative and synthetic positive controls are described in detail in \cite{suchard_comprehensive_2019}.  The number of effect size estimates that were estimated in this analysis for each data set are described in Tables~\ref{tab:SampleSizeTableDepression} and \ref{tab:SampleSizeTableHypertension}.

\begin{table}[htp]
\begin{center}
\begin{tabular}{lcccc}
\hline
 & \multicolumn{4}{c}{Effect Size} \\ 
Database & 1 & 1.5 & 2 & 4 \\ 
\hline
\rowcolor{Gray} CCAE  & $2617$ & $1957$ & $1957$ & $1957$ \\
MDCD  & $1725$ & $\phantom{0}836$ & $\phantom{0}836$ & $\phantom{0}836$ \\
\rowcolor{Gray} MDCR  & $1269$ & $\phantom{0}617$ & $\phantom{0}617$ & $\phantom{0}617$ \\
Optum  & $2415$ & $1805$ & $1806$ & $1806$ \\
\hline 
\end{tabular}
\end{center}
 \caption{Number of effect size estimates for the depression data set.}
    \label{tab:SampleSizeTableDepression}  
\end{table}

\begin{table}[htp]
\begin{center}
\begin{tabular}{lcccc}
\hline
 & \multicolumn{4}{c}{Effect Size} \\ 
Database & 1 & 1.5 & 2 & 4 \\ 
\hline
\rowcolor{Gray} CCAE  & $\phantom{0}52311$ & $\phantom{0}27197$ & $\phantom{0}27197$ & $\phantom{0}27197$ \\
CUMC  & $\phantom{00}1017$ & $\phantom{000}318$ & $\phantom{000}318$ & $\phantom{000}318$ \\
\rowcolor{Gray} IMSG  & $\phantom{00}3310$ & $\phantom{000}829$ & $\phantom{000}829$ & $\phantom{000}829$ \\
JMDC  & $\phantom{000}673$ & $\phantom{000}393$ & $\phantom{000}393$ & $\phantom{000}393$ \\
\rowcolor{Gray} MDCD  & $\phantom{00}7376$ & $\phantom{00}3292$ & $\phantom{00}3292$ & $\phantom{00}3292$ \\
MDCR  & $\phantom{0}11030$ & $\phantom{00}5809$ & $\phantom{00}5809$ & $\phantom{00}5809$ \\
\rowcolor{Gray} NHIS NSC  & $\phantom{00}1108$ & $\phantom{000}468$ & $\phantom{000}468$ & $\phantom{000}468$ \\
Optum  & $\phantom{0}49771$ & $\phantom{0}28234$ & $\phantom{0}28234$ & $\phantom{0}28232$ \\
\rowcolor{Gray} Panther  & $125064$ & $\phantom{0}38199$ & $\phantom{0}38199$ & $\phantom{0}38198$ \\
\hline 
\end{tabular}
\end{center}
 \caption{Number of effect size estimates for the hypertension data set.}
    \label{tab:SampleSizeTableHypertension}  
\end{table}

\section{Results}

\subsection{Depression}

\subsubsection{Constant Model of the True Bias}
We first consider the results for the constant model of the true bias on the depression data set.

\medskip
{\bf Train on Negative and Positive Controls}
\medskip

\begin{figure*}[!htbp]
\centering
\includegraphics[width = .9\linewidth]{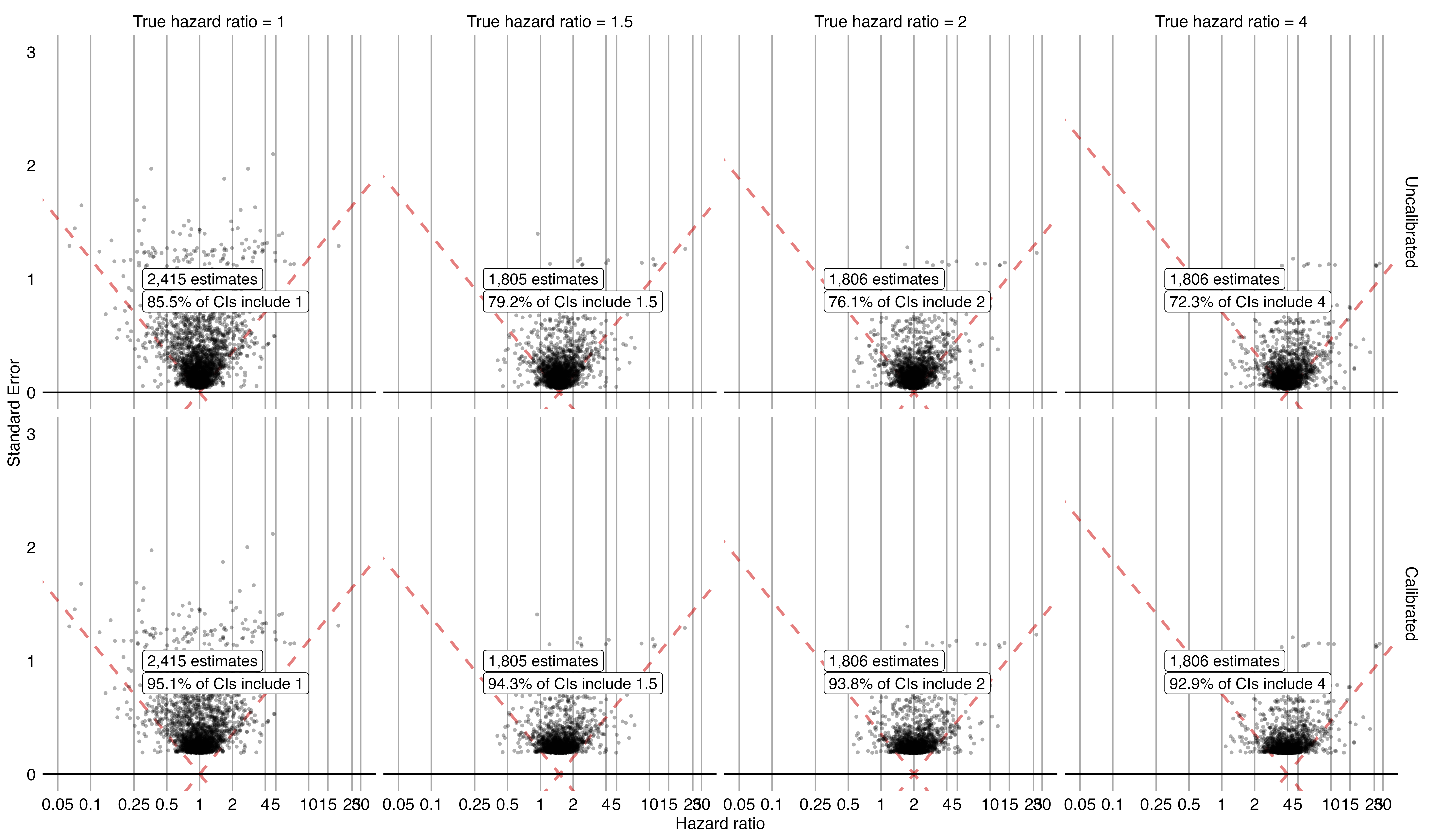}
\caption{Evaluation of the effect estimation before (top) and after (bottom) calibration using the Optum database in the depression data set.  The constant model of bias was used and training was performed on both negative and positive controls.}
\label{fig:Optum_AllControls_Depression}
\end{figure*}

We train on both negative and positive controls.  Figure \ref{fig:Optum_AllControls_Depression} shows the effect size estimates for all negative and positive controls as well as the percentage of posterior intervals containing the true effect size for the Optum database of the depression data set.  Each dot represents the hazard ratio and corresponding standard error for
one of the negative (true hazard ratio = 1) or positive control (true hazard ratio greater than 1) outcomes.  The coverage using the calibrated and uncalibrated methods for all of the databases in the depression data set is displayed in Table~\ref{tab:TableDepression}.

Pre-calibration posterior interval coverage departs significantly from 95\%, especially for the larger positive controls.  For all four databases, the calibrated procedure results in posterior intervals that are closer to the nominal 95\% coverage than the uncalibrated procedure.  

We also report the root mean squared error (RMSE) as a metric to quantify overall coverage across all estimates. We define RMSE as 

\begin{equation*}
RMSE = \sqrt{\frac{\sum_{t=1}^T(\hat{w} - w)^2}{T}}, 
\end{equation*}

where $\hat{w}$ is the number of times that the true effect size is contained in the confidence interval calculated by the calibrated and uncalibrated procedures, divided by the number of observations that were tested, $w = 0.95$  to represent the 95\% coverage of the true effect size, and $T = 4$ for the four hazard ratios (1, 1.5, 2, and 4).  

\begin{figure}[h!]
\centering
\includegraphics[width = 0.50\textwidth]{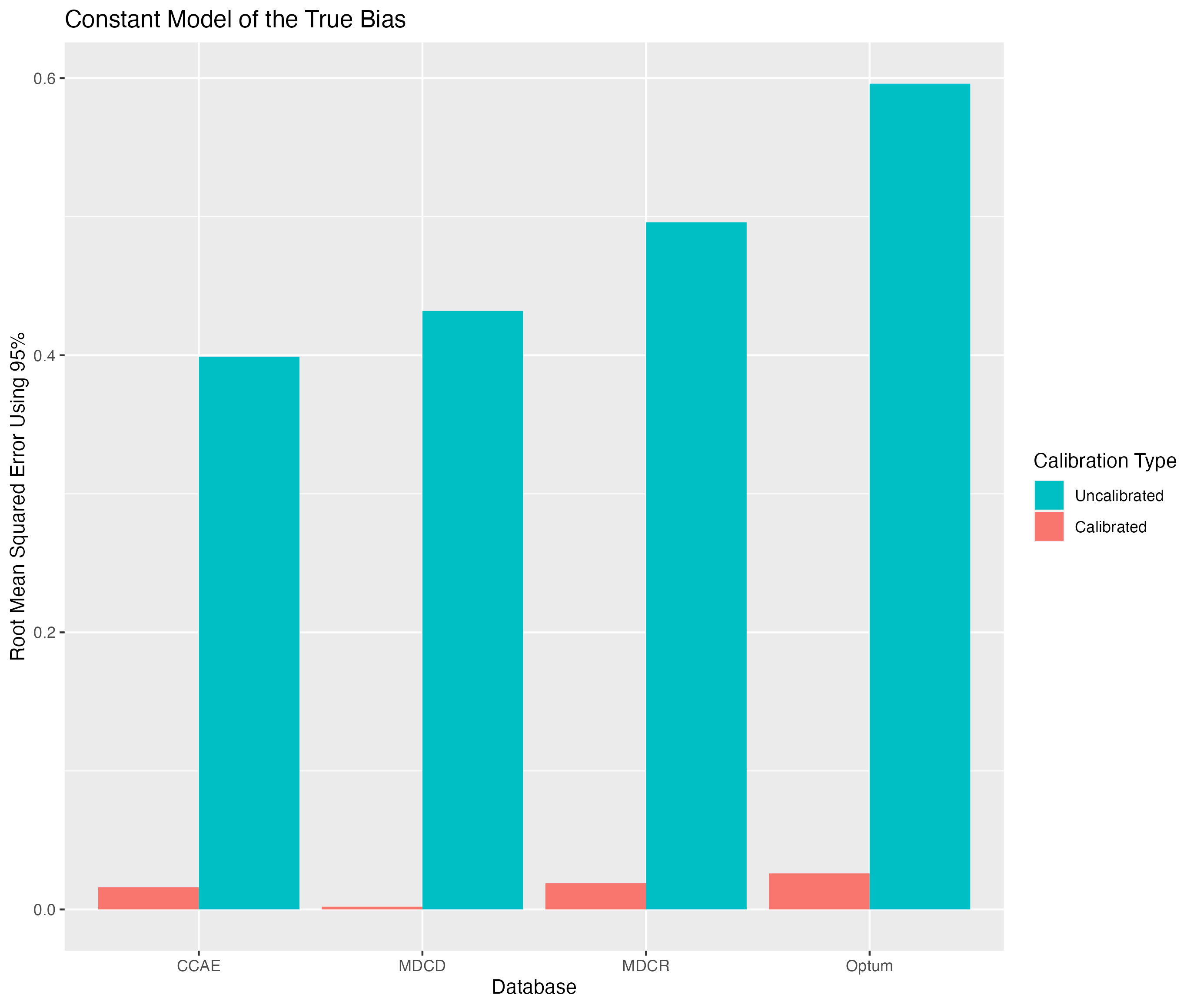}
\caption{Root mean squared error of all estimated effects before (uncalibrated) and after (calibrated) calibration for each of the databases in the depression data set. The constant model of bias was used and training was performed on both negative and positive controls.}
\label{fig:RMSE_AllControls_Depression}
\end{figure}

The RMSE for the Optum database using calibration is about 0.03 while without calibration, it is about 0.60.  Figure \ref{fig:RMSE_AllControls_Depression} shows the calculated RMSE for each database.   Overall, the uncalibrated procedure has higher error than the calibrated procedure.

\begin{table*}[htp]
\centering
\begin{tabular}{lllcccc}\hline
& & & \multicolumn{4}{c}{Effect Size} \\ 
& & &  1 & 1.5 & 2 & \multicolumn{1}{c}{4} \\ 
Training Type & Database & Calibration Type & Coverage & Coverage & Coverage & Coverage \\\hline
\rowcolor{Gray} Constant &  CCAE & Calibrated  & $\boldsymbol{0.947}$ & $\boldsymbol{0.941}$ & $\boldsymbol{0.945}$ & $\boldsymbol{0.935}$ \centering\arraybackslash\\
 &  & Uncalibrated  & $0.818$ & $0.758$ & $0.735$ & $0.691$ \\
\rowcolor{Gray}  & MDCD & Calibrated & $\boldsymbol{0.959}$ & $\boldsymbol{0.958}$ & $ \boldsymbol{0.953}$ &   $\boldsymbol{0.955}$\\
 &  & Uncalibrated  & $0.911$ & $0.850$ & $0.824$ & $0.792$ \\
\rowcolor{Gray} &  MDCR & Calibrated  & $\boldsymbol{0.953}$ & $\mathbf{0.938}$ & $\mathbf{0.929}$ & $\mathbf{0.922}$ \\
 &  & Uncalibrated  & $0.903$ & $0.830$ & $0.804$ & $0.767$ \\
\rowcolor{Gray} & Optum & Calibrated  & $\mathbf{0.951}$ & $\mathbf{0.943}$ & $ \mathbf{0.938}$ & $\mathbf{0.929}$ \\
 &  & Uncalibrated  & $0.855$ & $0.792$ & $0.761$ & $0.723$ \\
\rowcolor{Gray}  Linear & CCAE & Calibrated  & $\mathbf{0.946}$ & $\mathbf{0.942}$ & $\mathbf{0.944}$ & $\mathbf{0.945}$ \\
 &  & Uncalibrated  & $0.818$ & $0.758$ & $0.735$ & $0.691$ \\
\rowcolor{Gray}  &MDCD & Calibrated  &  $\mathbf{0.962}$ & $\mathbf{0.955}$ & $\mathbf{0.953}$ & $\mathbf{0.957}$ \\
 &  & Uncalibrated  & $0.911$ & $0.850$ & $0.824$ & $0.792$ \\
\rowcolor{Gray}  & MDCR & Calibrated  & $\mathbf{0.956}$ & $\mathbf{0.938}$ & $\mathbf{0.932}$ & $\mathbf{0.942}$ \\
 &  & Uncalibrated  & $0.903$ & $0.830$ & $0.804$ & $0.767$ \\
 \rowcolor{Gray} & Optum & Calibrated  & $ \mathbf{0.949}$ & $\mathbf{0.946}$ & $ \mathbf{0.941}$ & $ \mathbf{0.942}$ \\
 &  & Uncalibrated  & $0.855$ & $0.792$ & $0.761$ & $0.723$ \\
\rowcolor{Gray}  Train on Only Negative Controls &CCAE & Calibrated  & $\mathbf{0.942}$ & $ \mathbf{0.939}$ & $\mathbf{0.939}$ & $\mathbf{0.926}$ \\
 &  & Uncalibrated  & $0.818$ & $0.758$ & $0.735$ & $0.691$ \\
\rowcolor{Gray} &  MDCD & Calibrated  & $\mathbf{0.957}$ & $\mathbf{0.955}$ & $ \mathbf{0.951}$ & $\mathbf{0.947}$ \\
 &  & Uncalibrated  & $0.911$ & $0.850$ & $0.824$ & $0.792$ \\
 \rowcolor{Gray} &  MDCR & Calibrated  & $\mathbf{0.947}$ & $\mathbf{0.932}$ & $ \mathbf{0.916}$ & $\mathbf{0.900}$ \\
 &  & Uncalibrated  & $0.903$ & $0.830$ & $0.804$ & $0.767$ \\
 \rowcolor{Gray} &  Optum & Calibrated  & $\mathbf{0.944}$ & $\mathbf{0.936}$ & $ \mathbf{0.929}$ & $\mathbf{0.922}$ \\
 &  & Uncalibrated  & $0.855$ & $0.792$ & $0.761$ & $0.723$ \\\hline 
\end{tabular}
 \caption{Coverage for the calibrated and uncalibrated procedures on the depression data set.  Coverage closest to the 95th percentile is bolded.}
    \label{tab:TableDepression}  
\end{table*}

\medskip
{\bf Train on Only Negative Controls}
\medskip

\begin{figure*}[h!]
\centering
\includegraphics[width = 0.9\linewidth]{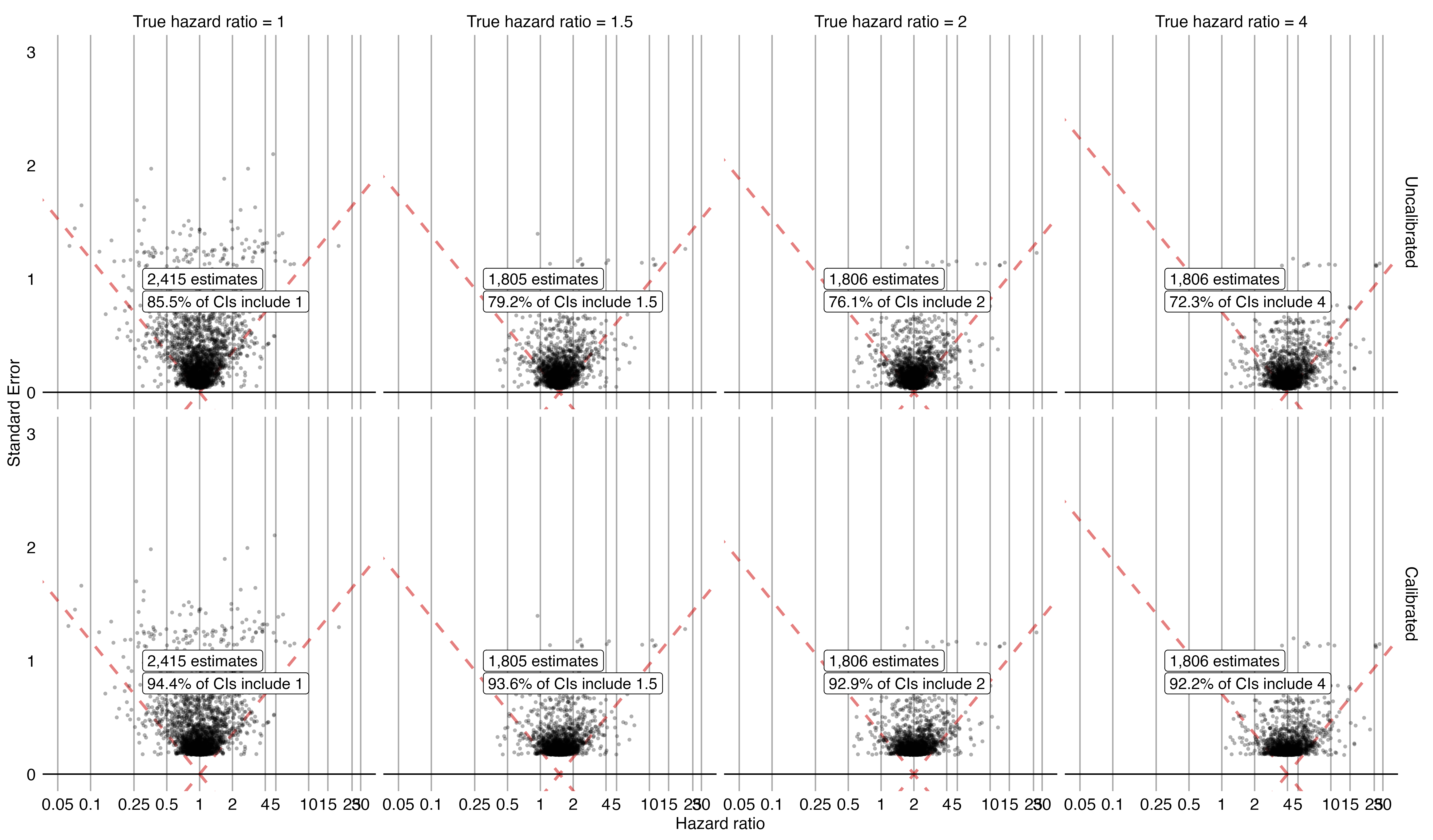}
\caption{Evaluation of the effect estimation after stratification on the propensity
scores before (top) and after (bottom) calibration using the Optum database in the depression data set.  The constant model of bias was used and training was performed on only negative controls.}
\label{fig:Optum_NegativeControls_Depression}
\end{figure*}

We also consider the effect of training on negative controls only.  Figure \ref{fig:Optum_NegativeControls_Depression} reports the effect size estimates for the Optum database for all negative and positive controls as well as the percentage of posterior intervals containing the true effect size.

\begin{figure*}[h!]
\centering
\includegraphics[width = 0.50\textwidth]{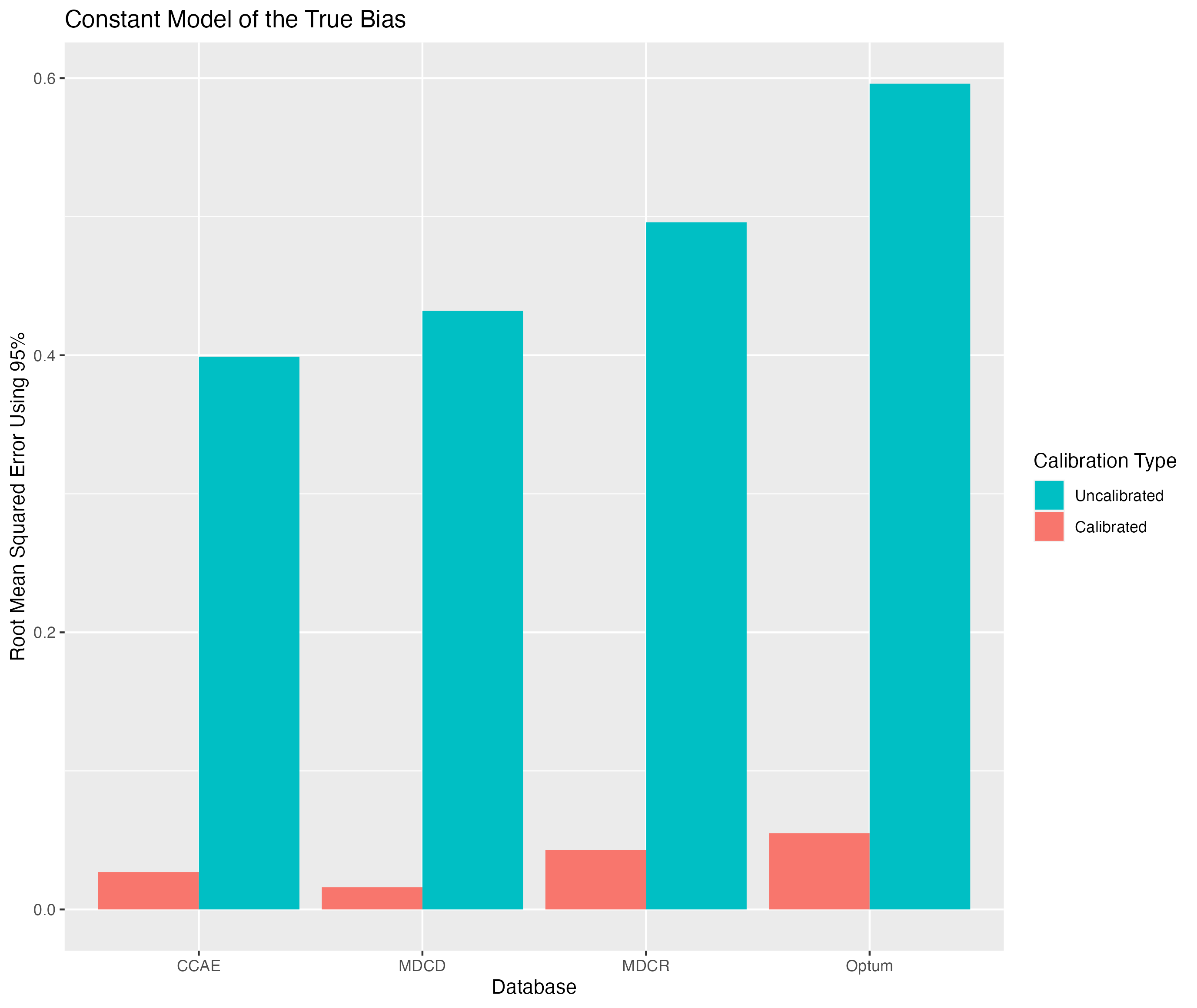}
\caption{Root mean squared error of all effects after stratification on the propensity
scores before (uncalibrated) and after (calibrated) calibration for each of the databases in the depression data set. The constant model of bias was used and training was performed on only negative controls.}
\label{fig:RMSE_NegativeControls_Depression}
\end{figure*}

The coverage for the calibrated and uncalibrated methods using the depression data set is displayed in Table~\ref{tab:TableDepression}.  For all four databases, the calibrated procedure results in posterior intervals that are closer to 95\% than the uncalibrated procedure.  When comparing the effect of training on both negative controls and positive controls to training only on negative controls for the calibrated procedure, overall, there appears to be a slight improvement in posterior interval coverage when training on both negative controls and positive controls.  The RMSE for the Optum database using calibration is about 0.06 while without calibration, it is about 0.60.  Figure \ref{fig:RMSE_NegativeControls_Depression} shows the RMSE calculated for each database.  Again, the uncalibrated procedure has higher error than the calibrated procedure, when considering all hazard ratios combined.  However, the RMSE for the calibrated procedure when training on only negative controls is higher, overall, than the RMSE for the calibrated procedure when trained on both negative and positive controls.

\subsubsection{Linear Model of the True Bias}
\begin{figure*}[h!]
\centering
\includegraphics[width = 0.9\linewidth]{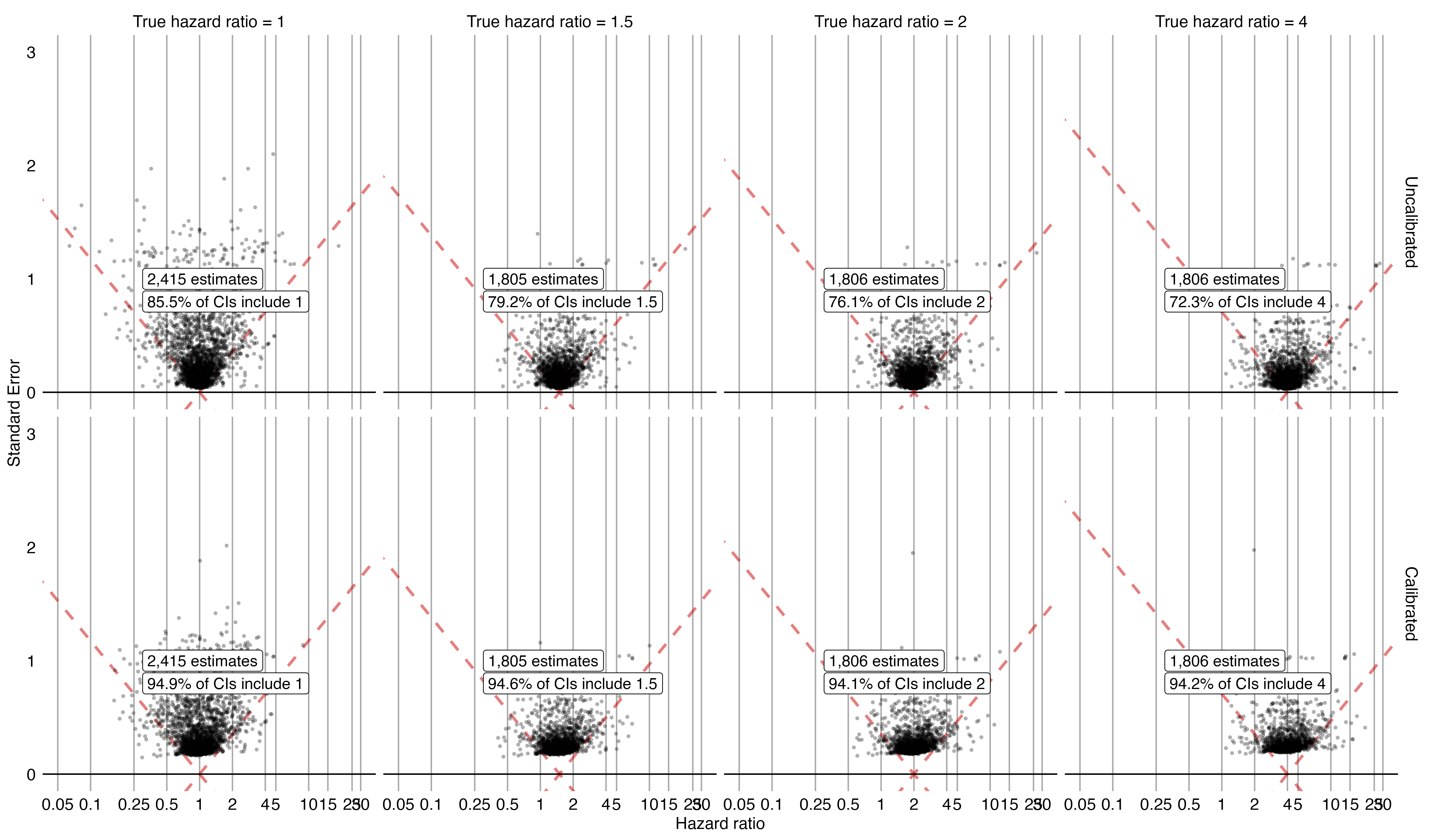}
\caption{Evaluation of the effect estimation before (top) and after (bottom) calibration using the Optum database in the depression data set.  The linear model of bias was used and training was performed on both negative and positive controls.}
\label{fig:Optum_Linear_Depression}
\end{figure*}

\begin{figure}[h!]
\centering
\includegraphics[width = 0.50\textwidth]{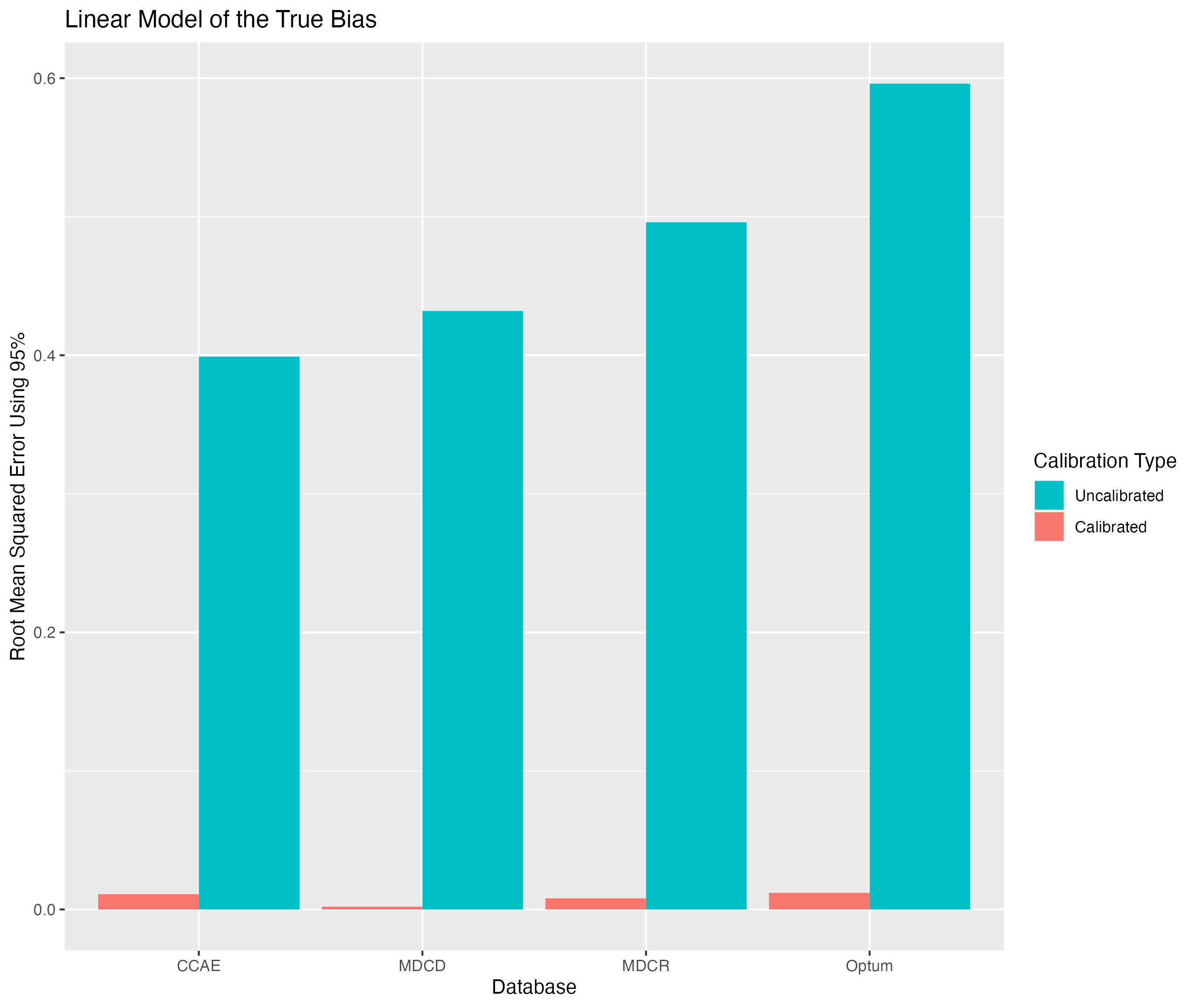}
\caption{Root mean squared error of all effects before (uncalibrated) and after (calibrated) calibration for each of the databases in the depression data set. The linear model of bias was used and training was performed on both negative and positive controls.}
\label{fig:RMSE_Linear_Depression}
\end{figure}

We review the results of the linear model of the true bias.  Figure \ref{fig:Optum_Linear_Depression} reports the effect size estimates for all negative and positive controls as well as the percentage of posterior intervals containing the true effect size for the Optum database of the hypertension data set.  The coverage for the calibrated and uncalibrated methods using the depression data set is displayed in Table~\ref{tab:TableDepression}.  For all four databases, the calibrated procedure results in posterior intervals that are closer to 95\% than the uncalibrated procedure.  The RMSE for the Optum database using calibration is about 0.01 while without calibration, it is about 0.60.  Figure \ref{fig:RMSE_Linear_Depression} shows the RMSE calculated for each database.  The uncalibrated procedure has higher error than the calibrated procedure, when considering all hazard ratios combined.  Out of the three approaches of training the data on the controls, the linear model of the true bias results in the lowest RMSE for all four databases of the depression data set.

\subsection{Hypertension}
We compare the calibrated confidence intervals using our proposed procedure to the confidence intervals generated by the gold standard, propensity score stratification.  We estimate 95\% posterior intervals and validate the proposed method on a data set for hypertension treatment. 
We use the same training and testing procedure, as well as the MCMC procedure, as used for the depression data set.   The trace, histogram, empirical cumulative distribution function, and autocorrelation plots all show convergence of the variables.

%We remove any missing relative risk estimates, standard errors, and true effect sizes.  Each observation in the data is a combination of a target, comparator, and outcome.  For the same target and comparator, one outcome is the negative control and three outcomes are the positive controls (1.5,2, and 4), unless no negative control outcome model is fitted and no positive controls are created due to the number of persons available.  As a result, the number of Target-Comparator-Outcome (TCO) observations for the CCAE database is 669928, 86420 for the MDCD database, 141499 for the MDCR database,  672425 for the Optum database, 9133 for the JMDC database, 29215 for the IMSG database, 9767 for the CUMC database, 1197225 for the Panther database, and  12265 for the NHIS NSC database. 

\subsubsection{Constant Model of the True Bias}
We first consider the results for the constant model of the true bias.

\medskip
{\bf Train on Negative and Positive Controls}
\medskip

We train on both negative and positive controls.  Figure \ref{fig:Optum_AllControls_Hypertension} shows the effect size estimates for all negative and positive controls as well as the percentage of posterior intervals containing the true effect size for the Optum database of the hypertension data set.   The coverage for the calibrated and uncalibrated methods using the hypertension data set for all nine databases is displayed in Table~\ref{tab:TableHypertension}. 

For Optum, CCAE, MDCD, MDCR, and Panther, the calibrated procedure results in posterior intervals that are closer to 95\% than the uncalibrated procedure.  However, for the other four databases, CUMC, JMDC, NHIS NSC, and IMSG, for the hazard ratio of one, the uncalibrated procedure performed slightly better than the calibrated procedure.  
Figure \ref{fig:RMSE_AllControls_Hypertension} shows the RMSE calculated for each database.  The uncalibrated procedure has higher error than the calibrated procedure, when considering all hazard ratios combined.

\medskip
{\bf Train on Only Negative Controls}
\medskip

We consider the effect of training on negative controls only.  Figure \ref{fig:Optum_NegativeControls_Hypertension} reports the effect size estimates for all negative and positive controls for the Optum database as well as the percentage of posterior intervals containing the true effect size.  The coverage for the calibrated and uncalibrated methods using the hypertension data set is displayed in Table~\ref{tab:TableHypertension}.  For Optum, CCAE, MDCD, MDCR, and Panther, the calibrated procedure results in posterior intervals that are closer to 95\% than the uncalibrated procedure.  The other four databases had similar results to the training on negative and positive controls analysis with the constant model of the true bias.   For the CUMC, JMDC, NHIS NSC, and IMSG databases, for the hazard ratio of one, the uncalibrated procedure performed slightly better than the calibrated procedure.  

 Figure \ref{fig:RMSE_NegativeControls_Hypertension} shows the RMSE calculated for each database.  Again, the uncalibrated procedure has higher error overall than the calibrated procedure, when considering all hazard ratios combined.  Similar to the depression data set, there appears to be more error overall when training on negative controls only compared to training on negative and positive controls.

\subsubsection{Linear Model of the True Bias}
We conclude with the results of the linear model of the true bias.  Figure \ref{fig:Optum_Linear_Hypertension} reports the effect size estimates for all negative and positive controls as well as the percentage of posterior intervals containing the true effect size for the Optum database of the hypertension data set.  The coverage for the calibrated and uncalibrated methods for all of the databases in the hypertension data set is displayed in Table~\ref{tab:TableHypertension}.  The performance of these databases is similar to the constant model of bias: 1) For Optum, CCAE, MDCD, MDCR, and Panther, the calibrated procedure results in posterior intervals that are closer to 95\% than the uncalibrated procedure,  2) For the CUMC, JMDC, NHIS NSC, and IMSG databases, for the hazard ratio of one, the uncalibrated procedure performed slightly better than the calibrated procedure, and 3) For JMDC, the uncalibrated method performed the best for the hazard ratio of 1.5.

Figure \ref{fig:RMSE_Linear_Hypertension} shows the RMSE calculated for each database.  The uncalibrated procedure has higher error overall than the calibrated procedure, when considering all hazard ratios combined.  Interestingly, for the calibrated procedure, the RMSE for the linear model of bias performed worse than the constant model of bias that was trained on both negative and positive controls for the hypertension data set, while the linear model performed better in terms of RMSE for the depression data set.  Thus, readers should evaluate different approaches in their specific contexts.

\subsection{Sensitivity Analysis}
We conducted a sensitivity analysis to assess the robustness of the results.  Firstly, we reviewed the uncertainty surrounding the hyperparameters $\mu, 1/\sigma^2, c,$ and $d$.  The mean and standard deviation of the prior parameters are included in the Appendix. Tables~\ref{tab:MeanStdDevPriors} and \ref{tab:MeanStdDevPriors_hypertension} have the statistics for the constant model using the depression and hypertension data sets, and Tables~\ref{tab:MeanStdDevPriors_linear_depression} and \ref{tab:MeanStdDevPriors_linear_hypertension} have the statistics for the linear model using the depression and hypertension data sets.  For the depression and hypertension data sets, the parameter $\mu$ is around zero with small variance.  Though the variance is small, the range of mean values for the parameter $1/\sigma^2$ is much wider between the two data sets.  For the depression data set, mean values for $1/\sigma^2$ are in the 20s to 30s.  The hypertension data set has a greater range in the mean values for $1/\sigma^2$ such that it could range from 5 to almost 100.  For the linear model of bias, the mean values for parameters $c$ and $d$ are around 0 with small variance.

Next, we looked at the lengths of the posterior intervals to assess whether the improved coverage is due to better centering, longer intervals, or a combination of both factors. 
Tables~\ref{tab:MeanStdDevPriors_intervallengths_depression}
and \ref{tab:MeanStdDevPriors_intervallengths_hypertension}.  It is evident that the improved coverage of the calibration method is generally due to longer intervals, which is expected to be wider after accounting for additional sources of bias. 

\begin{table*}[htp]
\centering
\scalebox{0.9}{
\begin{tabular}{lllcccc}
\hline
& & & \multicolumn{4}{c}{Effect Size} \\ 
& & & 1 & 1.5 & 2 & 4 \\ 
Training Type & Database & Calibration Type & Mean (Std Err) & Mean (Std Err) & Mean (Std Err) & Mean (Std Err)  \\ 
\hline
\rowcolor{Gray} Constant &  CCAE & Calibrated  & 1.657 ( 0.031 ) & 1.22 ( 0.014 ) & 1.199 ( 0.014 ) & 1.163 ( 0.014 ) \\
 &  & Uncalibrated  & 1.318 ( 0.033 ) & 0.799 ( 0.017 ) & 0.767 ( 0.017 ) & 0.709 ( 0.017 ) \\
 \rowcolor{Gray}  &MDCD & Calibrated  & 2.213 ( 0.032 ) & 1.269 ( 0.013 ) & 1.241 ( 0.013 ) & 1.195 ( 0.012 ) \\
 &  & Uncalibrated  & 1.956 ( 0.035 ) & 0.879 ( 0.017 ) & 0.839 ( 0.017 ) & 0.768 ( 0.016 ) \\
\rowcolor{Gray}  & MDCR & Calibrated  & 2.149 ( 0.038 ) & 1.166 ( 0.016 ) & 1.138 ( 0.016 ) & 1.091 ( 0.015 ) \\
 &  & Uncalibrated  & 1.929 ( 0.041 ) & 0.821 ( 0.02 ) & 0.781 ( 0.02 ) & 0.71 ( 0.019 ) \\
\rowcolor{Gray}  & Optum & Calibrated  & 1.662 ( 0.041 ) & 1.158 ( 0.013 ) & 1.164 ( 0.036 ) & 1.12 ( 0.035 ) \\
 &  & Uncalibrated  & 1.399 ( 0.043 ) & 0.829 ( 0.016 ) & 0.825 ( 0.037 ) & 0.761 ( 0.036 ) \\
\rowcolor{Gray} Linear &  CCAE & Calibrated  & 1.565 ( 0.018 ) & 1.198 ( 0.013 ) & 1.194 ( 0.013 ) & 1.19 ( 0.012 ) \\
 &  & Uncalibrated  & 1.318 ( 0.033 ) & 0.799 ( 0.017 ) & 0.767 ( 0.017 ) & 0.709 ( 0.017 ) \\
 \rowcolor{Gray}  &MDCD & Calibrated  & 2.109 ( 0.027 ) & 1.257 ( 0.013 ) & 1.239 ( 0.012 ) & 1.21 ( 0.012 ) \\
 &  & Uncalibrated  & 1.956 ( 0.035 ) & 0.879 ( 0.017 ) & 0.839 ( 0.017 ) & 0.768 ( 0.016 ) \\
\rowcolor{Gray}  & MDCR & Calibrated  & 2.029 ( 0.032 ) & 1.148 ( 0.016 ) & 1.141 ( 0.015 ) & 1.144 ( 0.014 ) \\
 &  & Uncalibrated  & 1.929 ( 0.041 ) & 0.821 ( 0.02 ) & 0.781 ( 0.02 ) & 0.71 ( 0.019 ) \\
  \rowcolor{Gray}  &Optum & Calibrated  & 1.525 ( 0.019 ) & 1.13 ( 0.012 ) & 1.127 ( 0.013 ) & 1.128 ( 0.012 ) \\
 &  & Uncalibrated  & 1.399 ( 0.043 ) & 0.829 ( 0.016 ) & 0.825 ( 0.037 ) & 0.761 ( 0.036 ) \\
\rowcolor{Gray}  Train on Only Negative Controls & CCAE & Calibrated  & 1.627 ( 0.031 ) & 1.186 ( 0.014 ) & 1.164 ( 0.014 ) & 1.128 ( 0.014 ) \\
 &  & Uncalibrated  & 1.318 ( 0.033 ) & 0.799 ( 0.017 ) & 0.767 ( 0.017 ) & 0.709 ( 0.017 ) \\
\rowcolor{Gray}  & MDCD & Calibrated  & 2.194 ( 0.032 ) & 1.242 ( 0.013 ) & 1.214 ( 0.013 ) & 1.166 ( 0.013 ) \\
 &  & Uncalibrated  & 1.956 ( 0.035 ) & 0.879 ( 0.017 ) & 0.839 ( 0.017 ) & 0.768 ( 0.016 ) \\
\rowcolor{Gray} &  MDCR & Calibrated  & 2.104 ( 0.038 ) & 1.099 ( 0.017 ) & 1.069 ( 0.016 ) & 1.019 ( 0.016 ) \\
 &  & Uncalibrated  & 1.929 ( 0.041 ) & 0.821 ( 0.02 ) & 0.781 ( 0.02 ) & 0.71 ( 0.019 ) \\
\rowcolor{Gray}  & Optum & Calibrated  & 1.621 ( 0.041 ) & 1.109 ( 0.014 ) & 1.116 ( 0.036 ) & 1.069 ( 0.036 ) \\
 &  & Uncalibrated  & 1.399 ( 0.043 ) & 0.829 ( 0.016 ) & 0.825 ( 0.037 ) & 0.761 ( 0.036 ) \\
\hline 
\end{tabular} }
 \caption{Table displaying mean and standard error of the posterior interval lengths for the depression data set.}
    \label{tab:MeanStdDevPriors_intervallengths_depression}  
\end{table*}

Finally, we checked to see how sensitive our results were to changes in the prior hyperparameters. We use uninformative priors in our original specification, where $\mu$ and $c$ are distributed $N(0,5)$ and $1/\sigma^2$ and $d$ are distributed $U(0,100)$.   We test with even more non-informative distributions, where $\mu$ and $c$ are both distributed as $N(0,10)$, and  $1/\sigma^2$ and $d$ are both distributed as $U(0,200)$.  We test the sensitivity of the results to the new specification using the depression data set and summarize the results in Table~\ref{tab:SensitivityAnalysisDepression} in the Appendix.  There is no material difference in the results that we obtained with our original specification and the new specification. Thus, the calibration method is not sensitive to the prior specification as long as the priors are not informative.

\section{Discussion}
In the context of two large-scale clinical studies, we have shown that calibrating posterior intervals restores interval coverage to near-nominal levels. This enables appropriate interpretation for decision making. Our results are similar to those of \cite{schuemie_empirical_2018}, but our Bayesian formulation provides a framework for further extension. For example, we can develop models that explicitly incorporate the biases and variance implicit in choice of database or in choice of analytic methods. Future work will take advantage of the Bayesian paradigm to enable Bayesian meta-analysis and, to account for model uncertainty, Bayesian model averaging.  We note that inclusion of synthetic positive controls impacted the results minimally. This may point to adopting a simpler approach of using only negative controls. Because they are simulated based on the negative controls, they do not add new information to the model, and further study of their benefit is warranted.

Our main assumptions are similar to those of \cite{schuemie_empirical_2018}.  We require that our negative controls are truly negative so that their effect sizes are zero. Our method is therefore dependent upon our process for generating negative controls, which generally entails considering exposure-outcome pairs to be negative when both the exposure and the outcome are well-studied but lack evidence that the exposure causes the outcome in scientific literature, structured product labels, or spontaneous adverse event reports. Lack of evidence of an effect is not the same as lack of an effect, but \cite{schuemie_empirical_2018} demonstrated the validity of this approach to negative controls (see its Appendix).

We also require that our controls are in some sense exchangeable with the intervention-outcome pairs of interest. We emphasize here that the bias need not be identical, but instead that the variety of types of confounding that exists among the controls is such that the confounding that affects the hypothesis of interest could have been drawn from a hypothetical distribution of confounding types seen in the controls.  We do not require the controls to have exactly the same magnitude and structure of confounding as the outcome of interest as other proposed approaches do \citep{tchetgen_tchetgen_control_2014, flanders_new_2017}.  For this study, we used the normal distribution to model the bias. Other distributions can be considered, such a $t$-distribution, or a Bayesian nonparametric approach. 

In summary, our Bayesian approach produced similar results to that of \cite{schuemie_empirical_2018} and provide a diagnostic to detect unmeasured confounding as well as a means to restore proper interval coverage at the expense of generally wider confidence intervals. Our approach complements the frequentist calibration and is amenable to extension to accommodate other sources of systematic error such as differences among databases and differences among analytic methods.

\section*{Acknowledgments}
This work was funded in part by National Institutes of Health grants R01 LM006910 and T15 LM007079.

\section*{Data Accessibility}
The data used in this article are freely available. Please contact the corresponding author for more information on how to obtain the data.

\begin{comment}
\subsection*{Author contributions}

This is an author contribution text. This is an author contribution text. This is an author contribution text. This is an author contribution text. This is an author contribution text. 
\end{comment}

\subsection*{Conflict of interest}

The authors declare no potential conflict of interests.

\begin{comment}
\section*{Supporting information}

The following supporting information is available as part of the online article:

\noindent
\textbf{Figure S1.}
{500{\uns}hPa geopotential anomalies for GC2C calculated against the ERA Interim reanalysis. The period is 1989--2008.}

\noindent
\textbf{Figure S2.}
{The SST anomalies for GC2C calculated against the observations (OIsst).}
\end{comment}

%\nocite{*}% Show all bib entries - both cited and uncited; comment this line to view only cited bib entries;
%\bibliography{wileyNJD-AMA}%
%\nocite{*}
\bibliography{Zotero.bib}
%\bibliography{WileyNJD-AMA}%
%\bibliographystyle{apa}
%\bibliographystyle{abbrvnat}
%\clearpage

\begin{comment}
\section*{Author Biography}

\begin{biography}{\includegraphics[width=66pt,height=86pt,draft]{empty}}{\textbf{Author Name.} This is sample author biography text this is sample author biography text this is sample author biography text this is sample author biography text this is sample author biography text this is sample author biography text this is sample author biography text this is sample author biography text this is sample author biography text this is sample author biography text this is sample author biography text this is sample author biography text this is sample author biography text this is sample author biography text this is sample author biography text this is sample author biography text this is sample author biography text this is sample author biography text this is sample author biography text this is sample author biography text this is sample author biography text.}
\end{biography}
\end{comment}

\appendix
\section{Additional Figures and Tables}

\begin{figure*}[htp]
\centering
\includegraphics[width = 0.9\linewidth]{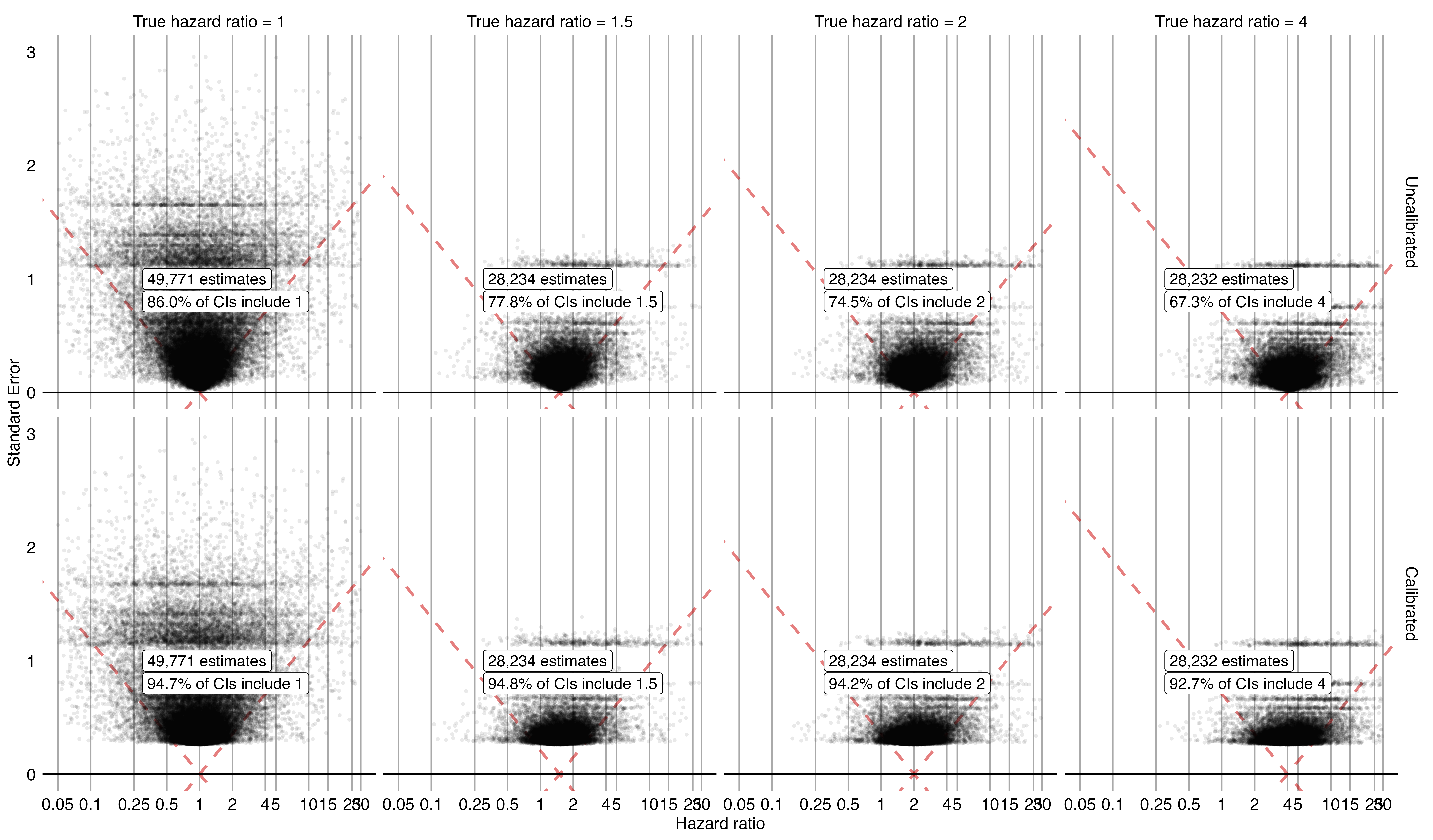}
\caption{Evaluation of the effect estimation before (top) and after (bottom) calibration using the Optum database in the hypertension data set. The constant model of bias was used and training was performed on both negative and positive controls.}
\label{fig:Optum_AllControls_Hypertension}
\end{figure*}

\begin{figure*}[htp]
\centering
\includegraphics[width = 0.50\textwidth]{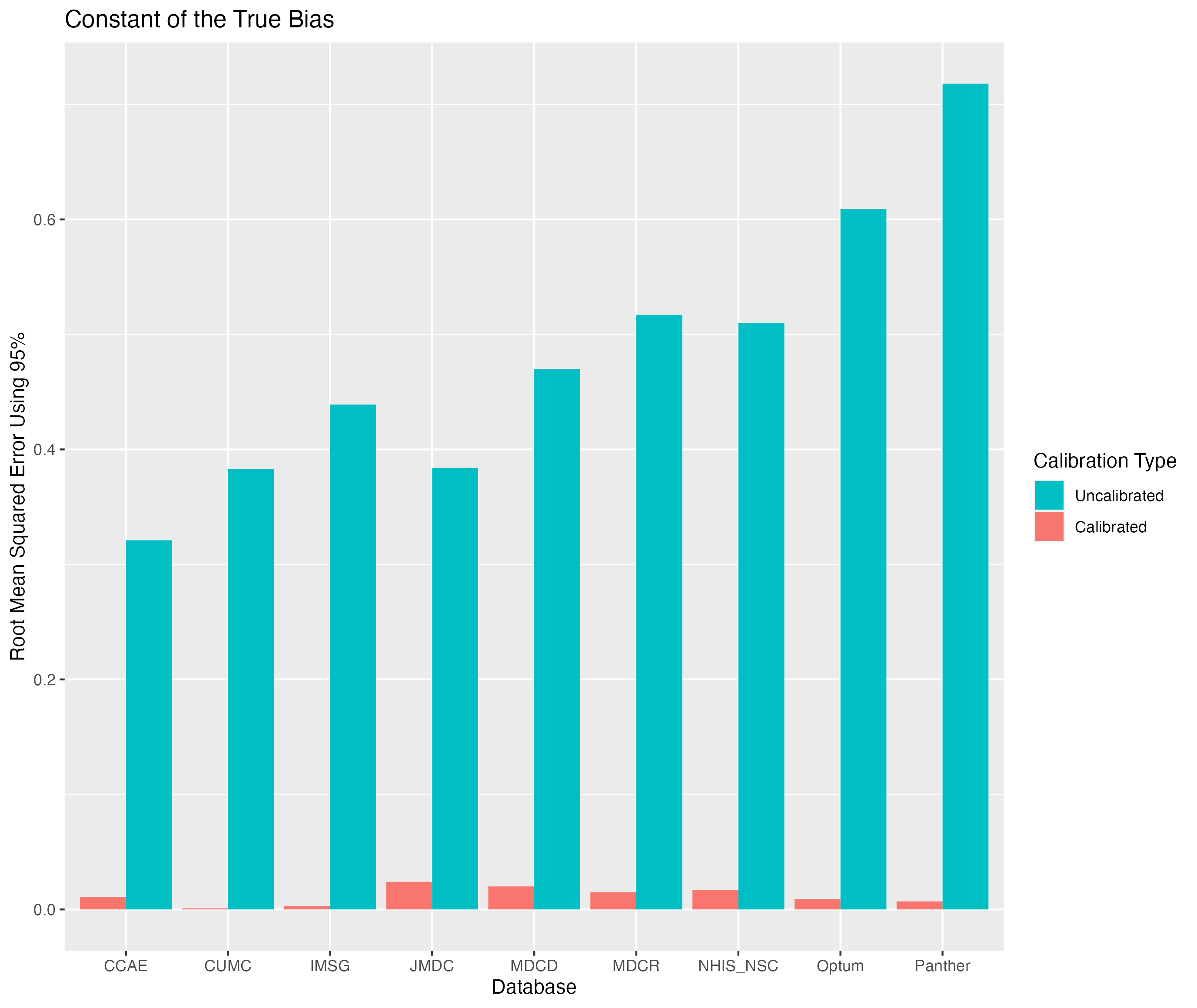}
\caption{Root mean squared error of all effects before (uncalibrated) and after (calibrated) calibration for each of the nine databases in the hypertension data set. The constant model of bias was used and training was performed on both negative and positive controls.}
\label{fig:RMSE_AllControls_Hypertension}
\end{figure*}

\begin{figure*}[h!]
\centering
\includegraphics[width = 0.9\linewidth]{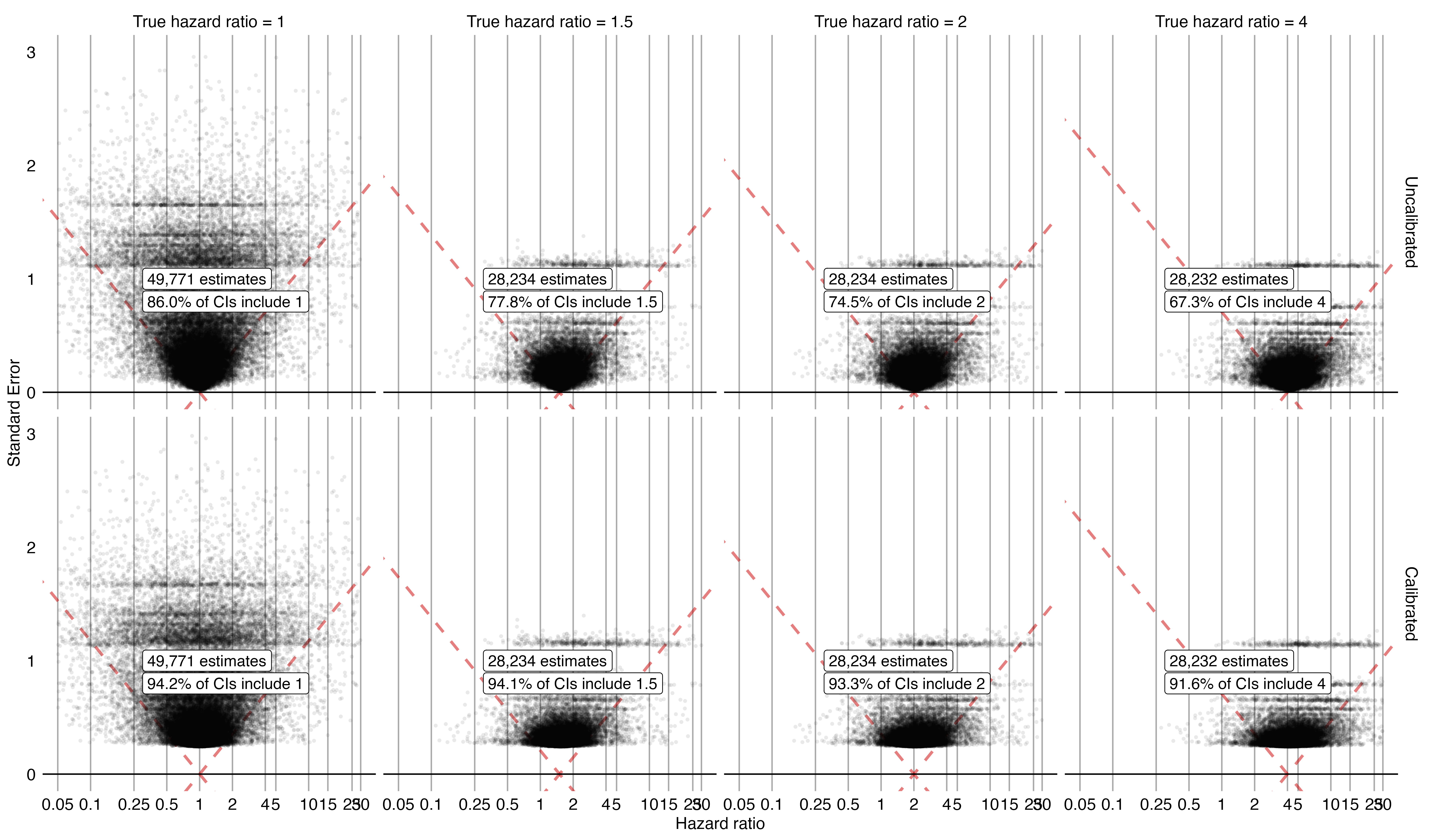}
\caption{Evaluation of the effect estimation before (top) and after (bottom) calibration using the Optum database in the hypertension data set.  The constant model of bias was used and training was performed on only negative controls.}
\label{fig:Optum_NegativeControls_Hypertension}
\end{figure*}

\begin{figure*}[h!]
\centering
\includegraphics[width = 0.50\textwidth]{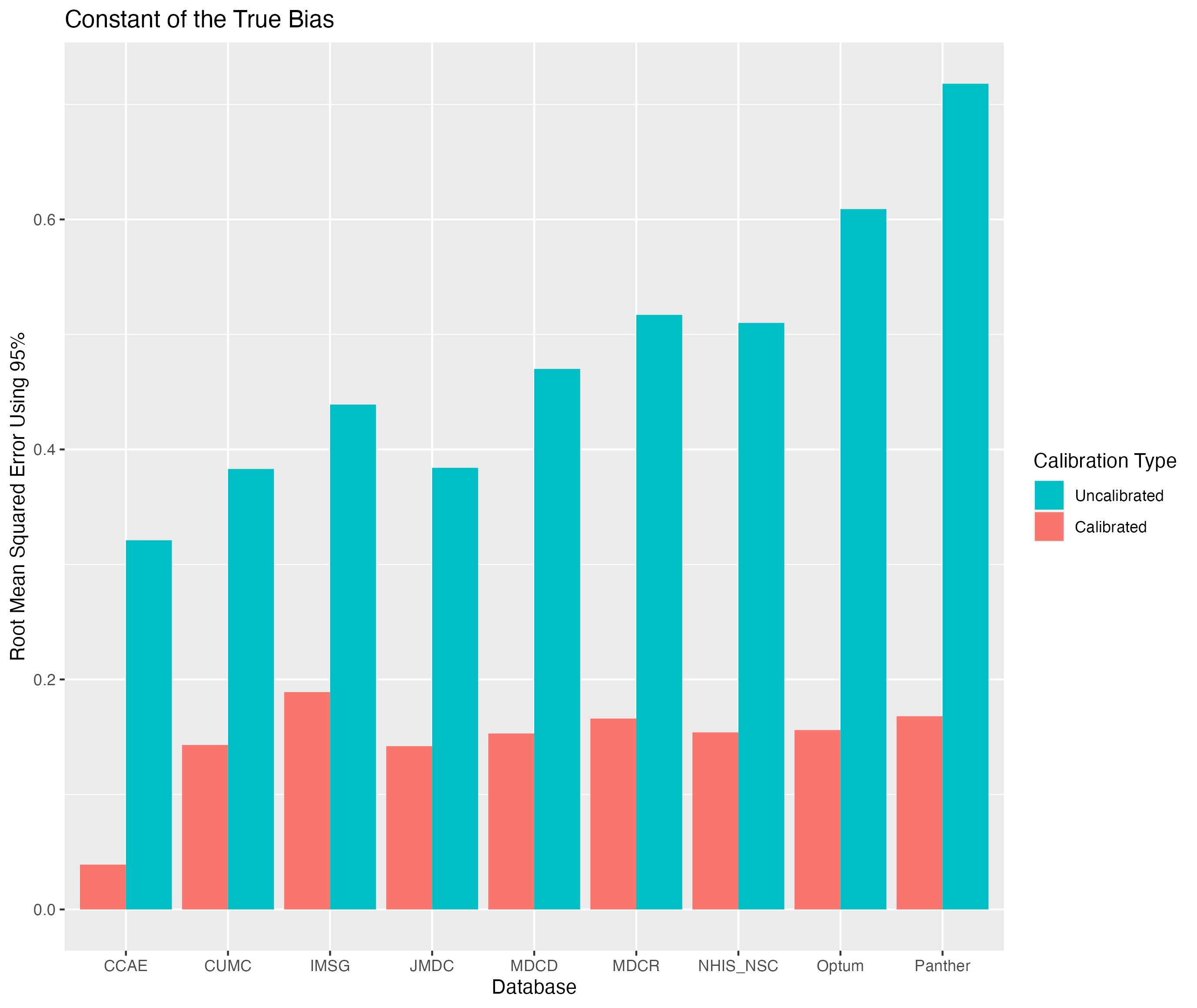}
\caption{Root mean squared error of all effects before (uncalibrated) and after (calibrated) calibration for each of the nine databases in the hypertension data set. The constant model of bias was used and training was performed on only negative controls. }
\label{fig:RMSE_NegativeControls_Hypertension}
\end{figure*}

\begin{table*}[htp]
\centering
\scalebox{0.9}{
\begin{tabular}{lllcccc}\hline
& & & \multicolumn{4}{c}{Effect Size} \\ 
& & & 1 & 1.5 & 2 & \multicolumn{1}{c}{4} \\ 
Training Type & Database & Calibration Type & Coverage & Coverage & Coverage & Coverage \\\hline
\rowcolor{Gray} Constant & CCAE & Calibrated  & $\mathbf{0.952}$ & $\mathbf{0.949}$ & $\mathbf{0.944}$ & $\mathbf{0.933}$ \\
 &  & Uncalibrated  & $0.877$ & $0.803$ & $0.773$ & $0.704$ \\
\rowcolor{Gray} & CUMC & Calibrated  & $0.978$ & $\mathbf{0.965}$ & $\mathbf{0.962}$ & $\mathbf{0.912}$ \\
 &  & Uncalibrated  & $\mathbf{0.966}$ & $0.852$ & $0.805$ & $0.736$ \\
 \rowcolor{Gray}& IMSG & Calibrated  & $0.978$ & $\mathbf{0.966}$ & $\mathbf{0.954}$ & $\mathbf{0.917}$ \\
 &  & Uncalibrated  & $\mathbf{0.953}$ & $0.858$ & $0.805$ & $0.748$ \\
\rowcolor{Gray} &  JMDC & Calibrated  & $0.975$ & $0.982$ & $\mathbf{0.967}$ & $\mathbf{0.962}$ \\
 &  & Uncalibrated  & $\mathbf{0.970}$ & $\mathbf{0.975}$ & $0.939$ & $0.901$ \\
 \rowcolor{Gray}  & MDCD & Calibrated  & $\mathbf{0.960}$ & $\mathbf{0.950}$ & $\mathbf{0.946}$ & $\mathbf{0.935}$ \\
 &  & Uncalibrated  & $0.898$ & $0.816$ & $0.786$ & $0.736$ \\
 \rowcolor{Gray}  & MDCR & Calibrated  & $\mathbf{0.962}$ & $\mathbf{0.954}$ & $\mathbf{0.944}$ & $\mathbf{0.924}$ \\
 &  & Uncalibrated  & $0.921$ & $0.856$ & $0.830$ & $0.761$ \\
 \rowcolor{Gray}&  NHIS NSC & Calibrated  & $0.968$ & $\mathbf{0.970}$ & $\mathbf{0.957}$ & $\mathbf{0.921}$ \\
 &  & Uncalibrated  & $\mathbf{0.951}$ & $0.929$ & $0.910$ & $0.844$ \\
 \rowcolor{Gray} &Optum & Calibrated  & $\mathbf{0.947}$ & $\mathbf{0.948}$ & $\mathbf{0.942}$ & $\mathbf{0.927}$ \\
 &  & Uncalibrated  & $0.860$ & $0.778$ & $0.745$ & $0.673$ \\
 \rowcolor{Gray} &Panther & Calibrated  & $\mathbf{0.954}$ & $\mathbf{0.960}$ & $\mathbf{0.948}$ & $\mathbf{0.925}$ \\
 &  & Uncalibrated  & $0.880$ & $0.755$ & $0.697$ & $0.603$ \\
 \rowcolor{Gray} Linear &  CCAE & Calibrated  & $\mathbf{0.956}$ & $\mathbf{0.949}$ & $\mathbf{0.948}$ & $\mathbf{0.949}$ \\
 &  & Uncalibrated  & $0.877$ & $0.803$ & $0.773$ & $0.704$ \\
 \rowcolor{Gray}& CUMC & Calibrated  & $0.980$ & $\mathbf{0.965}$ & $\mathbf{0.962}$ & $\mathbf{0.962}$ \\
 &  & Uncalibrated  & $\mathbf{0.966}$ & $0.852$ & $0.805$ & $0.736$ \\
\rowcolor{Gray} & IMSG & Calibrated  & $0.983$ & $\mathbf{0.959}$ & $\mathbf{0.960}$ & $\mathbf{0.972}$ \\
 &  & Uncalibrated  & $\mathbf{0.953}$ & $0.858$ & $0.805$ & $0.748$ \\
 \rowcolor{Gray}  & JMDC & Calibrated  & $0.976$ & $0.982$ & $\mathbf{0.967
}$ & $\mathbf{0.964}$ \\
 &  & Uncalibrated  & $\mathbf{0.970}$ & $\mathbf{0.975}$ & $0.901$ & $0.901$ \\
 \rowcolor{Gray} &  MDCD & Calibrated  & $\mathbf{0.964}$ & $\mathbf{0.948}$ & $\mathbf{0.953}$ & $\mathbf{0.962}$ \\
 &  & Uncalibrated  & $0.898$ & $0.816$ & $0.786$ & $0.736$ \\
 \rowcolor{Gray} &  MDCR & Calibrated  & $\mathbf{0.963}$ & $\mathbf{0.949}$ & $\mathbf{0.946}$ & $\mathbf{0.945}$ \\
 &  & Uncalibrated  & $0.921$ & $0.856$ & $0.830$ & $0.761$ \\
  \rowcolor{Gray} & NHIS NSC & Calibrated  & $0.977$ & $0.976$ & $\mathbf{0.964}$ & $\mathbf{0.932}$ \\
 &  & Uncalibrated  & $\mathbf{0.951}$ & $\mathbf{0.929}$ & $0.910$ & $0.844$ \\
 \rowcolor{Gray} &  Optum & Calibrated  & $\mathbf{0.949}$ & $\mathbf{0.949}$ & $\mathbf{0.946}$ & $\mathbf{0.942}$ \\
 &  & Uncalibrated  & $0.860$ & $0.778$ & $0.745$ & $0.673$ \\
 \rowcolor{Gray}&  Panther & Calibrated  & $\mathbf{0.958}$ & $\mathbf{0.956}$ & $\mathbf{0.951}$ & $\mathbf{0.948}$ \\
 &  & Uncalibrated  & $0.880$ & $0.755$ & $0.697$ & $0.603$ \\
\rowcolor{Gray} Train on Only Negative Controls  &  CCAE & Calibrated  & $\mathbf{0.943}$ & $\mathbf{0.937}$ & $\mathbf{0.930}$ & $\mathbf{0.913}$ \\
 &  & Uncalibrated  & $0.877$ & $0.803$ & $0.773$ & $0.704$ \\
 \rowcolor{Gray} & CUMC & Calibrated  & $0.970$ & $\mathbf{0.896}$ & $\mathbf{0.849}$ & $\mathbf{0.758}$ \\
 &  & Uncalibrated  & $\mathbf{0.966}$ & $0.852$ & $0.805$ & $0.736$ \\
 \rowcolor{Gray}&  IMSG & Calibrated  & $0.965$ & $\mathbf{0.908}$ & $\mathbf{0.865}$ & $\mathbf{0.813}$ \\
 &  & Uncalibrated  & $\mathbf{0.953}$ & $0.858$ & $0.805$ & $0.748$ \\
 \rowcolor{Gray}& JMDC & Calibrated  & $0.975$ & $0.980$ & $\mathbf{0.967}$ & $\mathbf{0.962}$ \\
 &  & Uncalibrated  & $\mathbf{0.970}$ & $\mathbf{0.975}$ & $0.939$ & $0.901$ \\
 \rowcolor{Gray}& MDCD & Calibrated  & $\mathbf{0.945}$ & $\mathbf{0.928}$ & $\mathbf{0.916}$ & $\mathbf{0.895}$ \\
 &  & Uncalibrated  & $0.898$ & $0.816$ & $0.786$ & $0.736$ \\
 \rowcolor{Gray}&  MDCR & Calibrated  & $\mathbf{0.949}$ & $\mathbf{0.931}$ & $\mathbf{0.915}$ & $\mathbf{0.879}$ \\
 &  & Uncalibrated  & $0.921$ & $0.856$ & $0.830$ & $0.761$ \\
 \rowcolor{Gray} & NHIS NSC & Calibrated  & $0.967$ & $\mathbf{0.966}$ & $\mathbf{0.957}$ & $\mathbf{0.908}$ \\
 &  & Uncalibrated  & $\mathbf{0.951}$ & $0.929$ & $0.910$ & $0.844$ \\
 \rowcolor{Gray} &Optum & Calibrated  & $\mathbf{0.942}$ & $\mathbf{0.941}$ & $\mathbf{0.933}$ & $0.916$ \\
 &  & Uncalibrated  & $0.860$ & $0.778$ & $0.745$ & $0.673$ \\
 \rowcolor{Gray}&  Panther & Calibrated  & $\mathbf{0.949}$ & $\mathbf{0.937}$ & $\mathbf{0.913}$ & $\mathbf{0.874}$ \\
 &  & Uncalibrated  & $0.880$ & $0.755$ & $0.697$ & $0.603$ \\\hline 
\end{tabular}}
 \caption{Table displaying coverage for the calibrated and uncalibrated procedures on the hypertension data set. Coverage closest to the 95th percentile is bolded.}
    \label{tab:TableHypertension}  
\end{table*}

\begin{figure*}[h!]
\centering
\includegraphics[width = 0.9\linewidth]{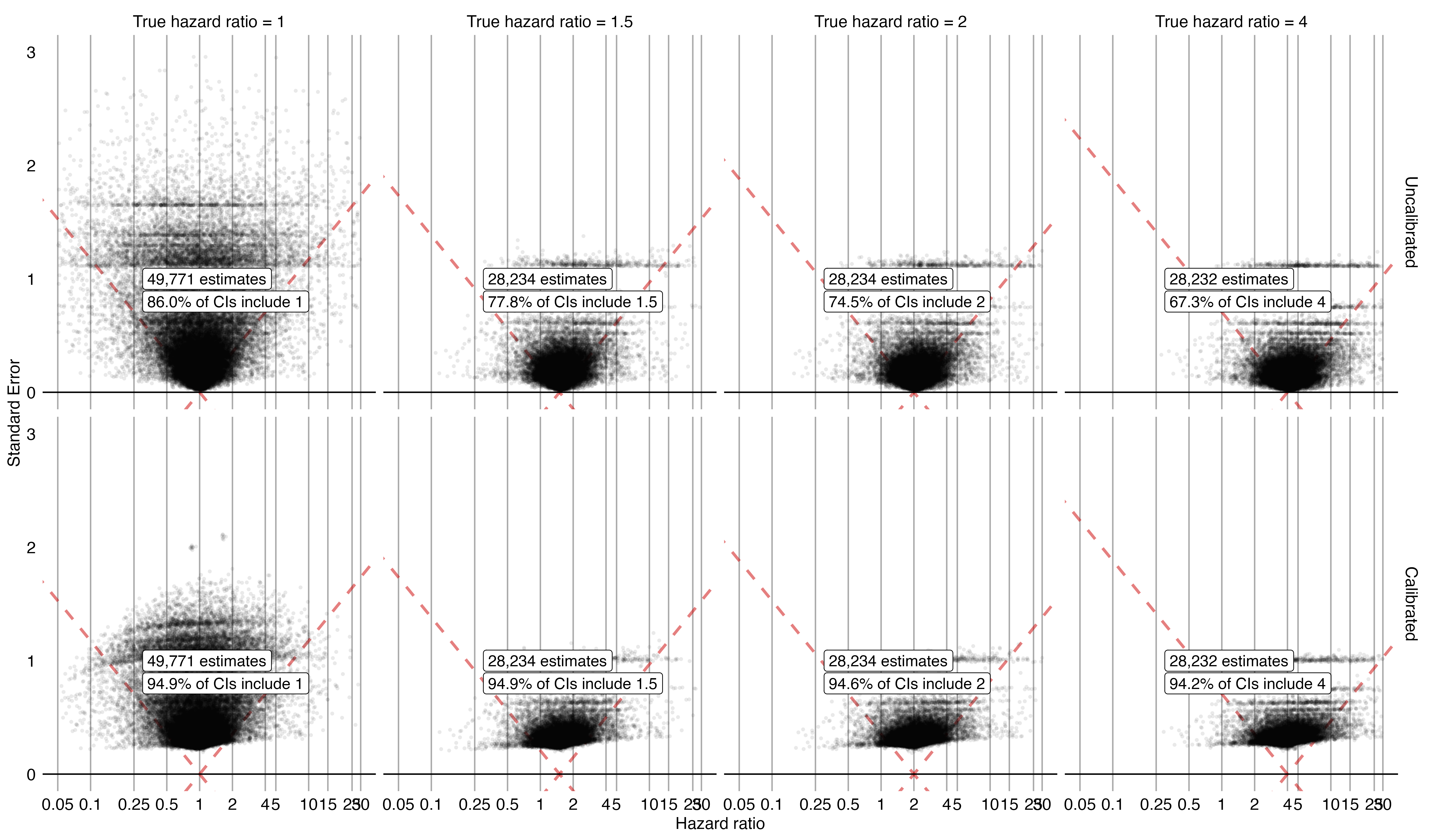}
\caption{Evaluation of the effect estimation before (top) and after (bottom) calibration using the Optum database in the hypertension data set.  The linear model of bias was used and training was performed on both negative and positive controls.}
\label{fig:Optum_Linear_Hypertension}
\end{figure*}

\begin{figure*}[h!]
\centering
\includegraphics[width = 0.50\textwidth]{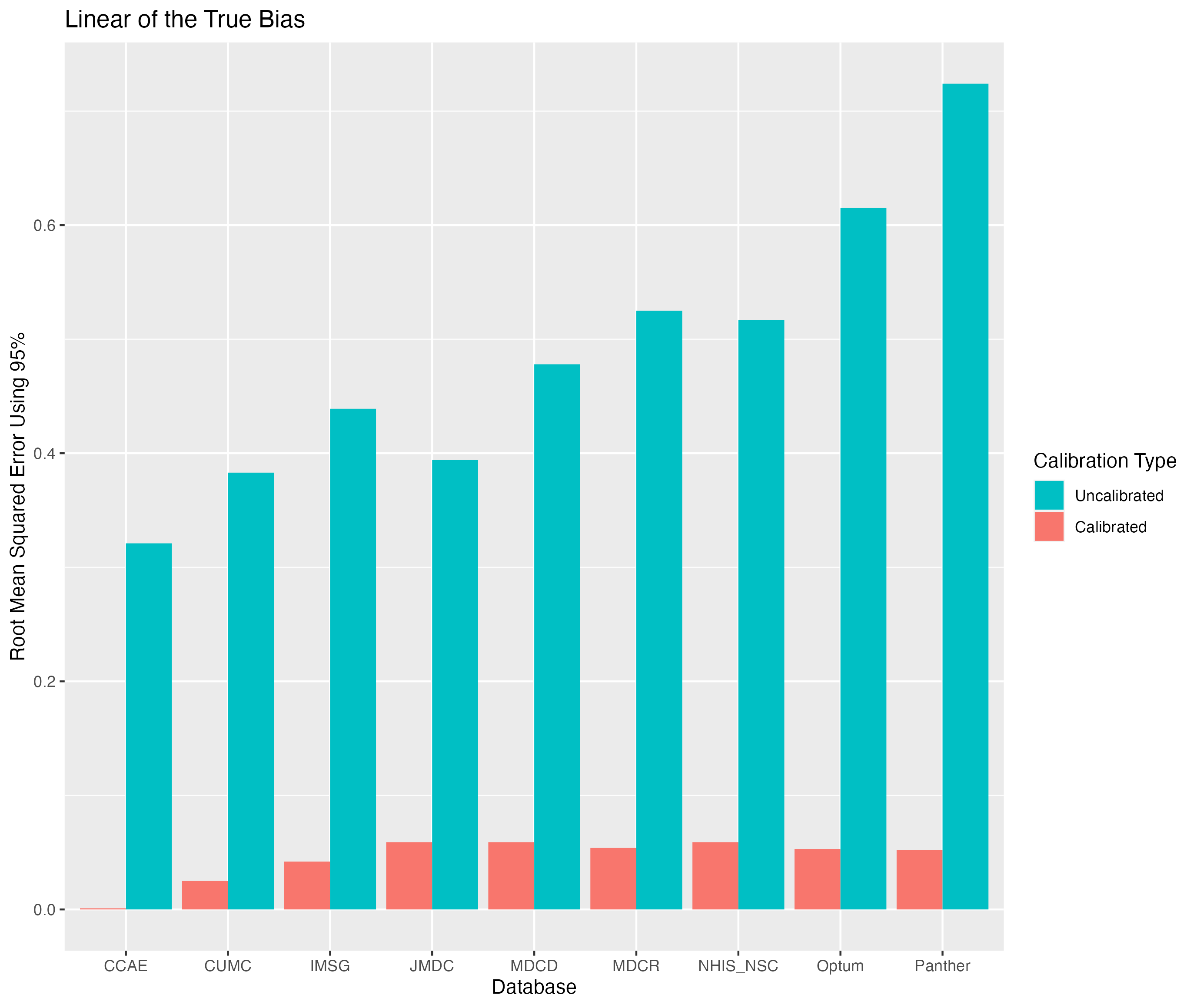}
\caption{Root mean squared error of all effects before (uncalibrated) and after (calibrated) calibration for each of the nine databases in the hypertension data set. The linear model of bias was used and training was performed on both negative and positive controls.}
\label{fig:RMSE_Linear_Hypertension}
\end{figure*}

\begin{table*}[htp]
\centering
\scalebox{0.9}{
\begin{tabular}{lllcccc}\hline
& & & \multicolumn{4}{c}{Effect Size} \\ 
& & & 1 & 1.5 & 2 & 4 \\ 
Training Type & Database & Prior Parameter & Mean (Std Dev) &Mean (Std Dev) & Mean (Std Dev)& Mean (Std Dev) \\\hline
\rowcolor{Gray} Constant &  CCAE & $\mu$  & 0.001 ( 0.001 ) & 0.001 ( 0.001 ) & 0.001 ( 0.001 ) & 0.001 ( 0.001 ) \\
 &  & $1/\sigma^2$  & 22.682 ( 0.267 ) & 22.684 ( 0.266 ) & 22.684 ( 0.27 ) & 22.682 ( 0.272 ) \\
 \rowcolor{Gray} & MDCD & $\mu$  & 0.009 ( 0.002 ) & 0.009 ( 0.002 ) & 0.009 ( 0.002 ) & 0.009 ( 0.002 ) \\
 &  & $1/\sigma^2$  & 20.442 ( 0.401 ) & 20.442 ( 0.396 ) & 20.438 ( 0.399 ) & 20.441 ( 0.4 ) \\
\rowcolor{Gray} &  MDCR & $\mu$  & 0.003 ( 0.003 ) & 0.003 ( 0.003 ) & 0.003 ( 0.003 ) & 0.003 ( 0.003 ) \\
 &  & $1/\sigma^2$  & 25.545 ( 0.613 ) & 25.564 ( 0.628 ) & 25.553 ( 0.635 ) & 25.554 ( 0.62 ) \\
\rowcolor{Gray}  & Optum & $\mu$  & 0.006 ( 0.001 ) & 0.005 ( 0.001 ) & 0.005 ( 0.001 ) & 0.005 ( 0.001 ) \\
 &  & $1/\sigma^2$  & 29.535 ( 0.432 ) & 29.531 ( 0.44 ) & 29.532 ( 0.427 ) & 29.528 ( 0.437 ) \\
\rowcolor{Gray} Train on Only Negative Controls &  CCAE & $\mu$  & 0 ( 0.003 ) & 0 ( 0.003 ) & 0 ( 0.003 ) & 0 ( 0.003 ) \\
 &  & $1/\sigma^2$  & 25.434 ( 0.643 ) & 25.441 ( 0.651 ) & 25.45 ( 0.64 ) & 25.443 ( 0.643 ) \\
 \rowcolor{Gray}  & MDCD & $\mu$  & 0.003 ( 0.005 ) & 0.003 ( 0.005 ) & 0.003 ( 0.005 ) & 0.003 ( 0.005 ) \\
 &  & $1/\sigma^2$  & 22.371 ( 0.936 ) & 22.374 ( 0.957 ) & 22.364 ( 0.973 ) & 22.351 ( 0.935 ) \\
\rowcolor{Gray}  & MDCR & $\mu$  & -0.002 ( 0.005 ) & -0.002 ( 0.005 ) & -0.002 ( 0.005 ) & -0.002 ( 0.005 ) \\
 &  & $1/\sigma^2$  & 33.342 ( 1.874 ) & 33.339 ( 1.88 ) & 33.376 ( 1.871 ) & 33.396 ( 1.885 ) \\
\rowcolor{Gray} &  Optum & $\mu$  & 0.001 ( 0.003 ) & 0.001 ( 0.003 ) & 0.001 ( 0.003 ) & 0.001 ( 0.003 ) \\
 &  & $1/\sigma^2$  & 36.325 ( 1.223 ) & 36.333 ( 1.225 ) & 36.324 ( 1.221 ) & 36.36 ( 1.226 ) \\\hline 
\end{tabular}}
 \caption{Table displaying mean and standard deviations of the prior parameters for the depression data set.}
    \label{tab:MeanStdDevPriors}  
\end{table*}

\begin{table*}[htp]
\centering
\scalebox{0.9}{
\begin{tabular}{lllcccc}
\hline
& & & \multicolumn{4}{c}{Effect Size} \\ 
& & & 1 & 1.5 & 2 & 4 \\ 
Training Type & Database & Prior Parameter & Mean (Std Dev) &  Mean (Std Dev)  &  Mean (Std Dev)  &  Mean (Std Dev)  \\ \hline
\rowcolor{Gray} Constant &  CCAE & $\mu$  & 0.042 ( 0.001 ) & 0.042 ( 0.001 ) & 0.042 ( 0.001 ) & 0.042 ( 0.001 ) \\
 &  & $1/\sigma^2$  & 9.508 ( 0.029 ) & 9.507 ( 0.029 ) & 9.507 ( 0.029 ) & 9.508 ( 0.029 ) \\
 \rowcolor{Gray} &  CUMC & $\mu$  & 0.114 ( 0.004 ) & 0.114 ( 0.004 ) & 0.114 ( 0.004 ) & 0.114 ( 0.004 ) \\
 &  & $1/\sigma^2$   & 22.367 ( 0.92 ) & 22.381 ( 0.929 ) & 22.372 ( 0.936 ) & 22.378 ( 0.947 ) \\
 \rowcolor{Gray} &  IMSG & $\mu$  & 0.179 ( 0.004 ) & 0.179 ( 0.004 ) & 0.179 ( 0.004 ) & 0.179 ( 0.004 ) \\
 &  & $1/\sigma^2 $  & 7.488 ( 0.151 ) & 7.491 ( 0.15 ) & 7.49 ( 0.15 ) & 7.49 ( 0.151 ) \\
 \rowcolor{Gray}  & JMDC & $\mu$  & 0.018 ( 0.002 ) & 0.018 ( 0.002 ) & 0.018 ( 0.002 ) & 0.018 ( 0.002 ) \\
 &  & $1/\sigma^2$   & 99.589 ( 0.401 ) & 99.582 ( 0.41 ) & 99.599 ( 0.391 ) & 99.578 ( 0.416 ) \\
 \rowcolor{Gray} &  MDCD & $\mu$  & 0.064 ( 0.002 ) & 0.064 ( 0.002 ) & 0.064 ( 0.002 ) & 0.064 ( 0.002 ) \\
 &  & $1/\sigma^2 $  & 9.179 ( 0.082 ) & 9.181 ( 0.084 ) & 9.181 ( 0.083 ) & 9.179 ( 0.084 ) \\
 \rowcolor{Gray}  & MDCR & $\mu$  & 0.023 ( 0.001 ) & 0.023 ( 0.001 ) & 0.023 ( 0.001 ) & 0.023 ( 0.001 ) \\
 &  & $1/\sigma^2$   & 29.714 ( 0.274 ) & 29.712 ( 0.272 ) & 29.718 ( 0.275 ) & 29.718 ( 0.272 ) \\
 \rowcolor{Gray}  & NHIS NSC & $\mu$  & 0.026 ( 0.003 ) & 0.026 ( 0.003 ) & 0.026 ( 0.003 ) & 0.026 ( 0.002 ) \\
 &  & $1/\sigma^2 $  & 98.266 ( 1.614 ) & 98.322 ( 1.55 ) & 98.347 ( 1.538 ) & 98.345 ( 1.526 ) \\
 \rowcolor{Gray} &  Optum & $\mu $ & 0.03 ( 0 ) & 0.03 ( 0 ) & 0.03 ( 0 ) & 0.03 ( 0 ) \\
 &  & $1/\sigma^2$   & 14.453 ( 0.047 ) & 14.453 ( 0.048 ) & 14.453 ( 0.046 ) & 14.454 ( 0.048 ) \\
 \rowcolor{Gray} & Panther & $\mu$  & 0.16 ( 0.001 ) & 0.16 ( 0.001 ) & 0.16 ( 0.001 ) & 0.16 ( 0.001 ) \\
 &  & $1/\sigma^2 $  & 5.301 ( 0.013 ) & 5.301 ( 0.013 ) & 5.301 ( 0.013 ) & 5.301 ( 0.013 ) \\
\rowcolor{Gray} Train on Only Negative Controls &  CCAE & $\mu$  & 0 ( 0.001 ) & 0 ( 0.001 ) & 0 ( 0.001 ) & 0 ( 0.001 ) \\
 &  & $1/\sigma^2$   & 13.47 ( 0.092 ) & 13.47 ( 0.094 ) & 13.472 ( 0.094 ) & 13.47 ( 0.094 ) \\
 \rowcolor{Gray}  & CUMC & $\mu$ & -0.002 ( 0.007 ) & -0.002 ( 0.007 ) & -0.002 ( 0.007 ) & -0.002 ( 0.007 ) \\
 &  & $1/\sigma^2 $  & 93.82 ( 5.371 ) & 93.82 ( 5.302 ) & 93.564 ( 5.557 ) & 93.789 ( 5.289 ) \\
 \rowcolor{Gray}  & IMSG & $\mu$  & 0.003 ( 0.005 ) & 0.003 ( 0.005 ) & 0.003 ( 0.005 ) & 0.003 ( 0.005 ) \\
 &  & $1/\sigma^2$   & 32.139 ( 2.945 ) & 32.187 ( 2.888 ) & 32.147 ( 2.884 ) & 32.125 ( 2.889 ) \\
 \rowcolor{Gray}  & JMDC & $\mu$  & 0.001 ( 0.005 ) & 0.001 ( 0.005 ) & 0.001 ( 0.005 ) & 0.001 ( 0.005 ) \\
 &  & $1/\sigma^2 $  & 98.628 ( 1.316 ) & 98.669 ( 1.303 ) & 98.64 ( 1.309 ) & 98.617 ( 1.311 ) \\
 \rowcolor{Gray}&  MDCD & $\mu$  & 0 ( 0.003 ) & 0 ( 0.003 ) & 0 ( 0.003 ) & 0 ( 0.003 ) \\
 &  & $1/\sigma^2$   & 16.796 ( 0.385 ) & 16.794 ( 0.383 ) & 16.793 ( 0.377 ) & 16.788 ( 0.388 ) \\
 \rowcolor{Gray} & MDCR & $\mu$  & 0 ( 0.001 ) & 0 ( 0.001 ) & 0 ( 0.001 ) & 0 ( 0.001 ) \\
 &  & $1/\sigma^2$   & 57.23 ( 1.586 ) & 57.219 ( 1.618 ) & 57.223 ( 1.559 ) & 57.187 ( 1.571 ) \\
 \rowcolor{Gray}&  NHIS NSC & $\mu$ & -0.002 ( 0.005 ) & -0.002 ( 0.005 ) & -0.002 ( 0.005 ) & -0.002 ( 0.005 ) \\
 &  & $1/\sigma^2$   & 97.571 ( 2.295 ) & 97.577 ( 2.214 ) & 97.622 ( 2.262 ) & 97.537 ( 2.348 ) \\
 \rowcolor{Gray}&  Optum & $\mu $ & 0 ( 0.001 ) & 0 ( 0.001 ) & 0 ( 0.001 ) & 0 ( 0.001 ) \\
 &  & $1/\sigma^2$   & 16.882 ( 0.122 ) & 16.883 ( 0.12 ) & 16.883 ( 0.12 ) & 16.879 ( 0.122 ) \\
 \rowcolor{Gray} &  Panther & $\mu$  & 0 ( 0.001 ) & 0 ( 0.001 ) & 0 ( 0.001 ) & 0 ( 0.001 ) \\
 &  & $1/\sigma^2$   & 7.09 ( 0.034 ) & 7.091 ( 0.034 ) & 7.09 ( 0.034 ) & 7.09 ( 0.034 ) \\\hline 
\end{tabular}}
 \caption{Table displaying mean and standard deviations of the prior parameters for the hypertension data set.}
    \label{tab:MeanStdDevPriors_hypertension}  
\end{table*}

\begin{table*}[htp]
\centering
\begin{tabular}{lllcccc}
\hline
& & & \multicolumn{4}{c}{Effect Size} \\ 
& & & 1 & 1.5 & 2 & 4 \\ 
Training Type & Database & Prior Parameter & Mean (Std Dev) & Mean (Std Dev)  & Mean (Std Dev)  & Mean (Std Dev) \\ 
\hline
\rowcolor{Gray} Linear &  CCAE & c  & -0.02 ( 0.003 ) & -0.016 ( 0.003 ) & -0.015 ( 0.003 ) & -0.012 ( 0.003 )\\
 &  & d  & 0.008 ( 0.001 ) & 0.008 ( 0.001 ) & 0.008 ( 0.001 ) & 0.007 ( 0.001 ) \\
\rowcolor{Gray}  &  & $\mu$  & 0.013 ( 0.002 ) & 0.011 ( 0.002 ) & 0.011 ( 0.002 ) & 0.009 ( 0.002 ) \\
 &  & $1/\sigma^2$  & 25.678 ( 0.518 ) & 25.622 ( 0.518 ) & 25.523 ( 0.507 ) & 25.424 ( 0.514 ) \\
\rowcolor{Gray} &  MDCD & c  & -0.029 ( 0.005 ) & -0.012 ( 0.005 ) & -0.012 ( 0.005 ) & -0.008 ( 0.005 ) \\
 &  & d  & 0.006 ( 0.002 ) & 0.006 ( 0.002 ) & 0.005 ( 0.002 ) & 0.005 ( 0.002 ) \\
 \rowcolor{Gray} &  & $\mu$  & 0.026 ( 0.004 ) & 0.017 ( 0.004 ) & 0.017 ( 0.004 ) & 0.015 ( 0.004 ) \\
 &  & $1/\sigma^2$  & 22.281 ( 0.719 ) & 22.153 ( 0.74 ) & 22.098 ( 0.745 ) & 21.904 ( 0.724 ) \\
\rowcolor{Gray} &  MDCR & c  & -0.028 ( 0.005 ) & -0.014 ( 0.005 ) & -0.013 ( 0.005 ) & -0.009 ( 0.005 ) \\
 &  & d  & 0.009 ( 0.001 ) & 0.011 ( 0.002 ) & 0.011 ( 0.002 ) & 0.011 ( 0.002 ) \\
\rowcolor{Gray} &   & $\mu$  & 0.02 ( 0.004 ) & 0.012 ( 0.004 ) & 0.012 ( 0.004 ) & 0.01 ( 0.004 ) \\
 &  & $1/\sigma^2$  & 29.979 ( 0.997 ) & 31.771 ( 1.47 ) & 31.729 ( 1.459 ) & 31.575 ( 1.403 ) \\
\rowcolor{Gray}  & Optum & c  & -0.014 ( 0.003 ) & -0.01 ( 0.003 ) & -0.009 ( 0.003 ) & -0.006 ( 0.003 ) \\
 &  & d  & 0.009 ( 0.001 ) & 0.01 ( 0.001 ) & 0.009 ( 0.001 ) & 0.009 ( 0.001 ) \\
\rowcolor{Gray}  &  & $\mu$  & 0.014 ( 0.002 ) & 0.012 ( 0.002 ) & 0.012 ( 0.002 ) & 0.01 ( 0.002 ) \\
 &  & $1/\sigma^2$  & 35.514 ( 0.779 ) & 36.413 ( 1 ) & 36.324 ( 0.961 ) & 36.105 ( 0.986 ) \\
\hline 
\end{tabular} 
 \caption{Table displaying mean and standard deviations of the prior parameters for the depression data set for the linear model of bias.}
    \label{tab:MeanStdDevPriors_linear_depression}  
\end{table*}

\begin{table*}[htp]
\centering
\begin{tabular}{lllcccc}
\hline
& & & \multicolumn{4}{c}{Effect Size} \\ 
& & & 1 & 1.5 & 2 & 4 \\ 
Training Type & Database & Prior Parameter & Mean (Std Dev) & Mean (Std Dev) & Mean (Std Dev) & Mean (Std Dev) \\ 
\hline
\rowcolor{Gray} Linear &  CCAE & c  & -0.001 ( 0.002 ) & 0.023 ( 0.001 ) & 0.024 ( 0.001 ) & 0.032 ( 0.001 ) \\
 &  & d  & 0.029 ( 0.003 ) & 0.041 ( 0.001 ) & 0.042 ( 0.001 ) & 0.043 ( 0.001 ) \\
 \rowcolor{Gray}&   & $\mu$  & 0.04 ( 0.001 ) & 0.027 ( 0.001 ) & 0.027 ( 0.001 ) & 0.024 ( 0.001 ) \\
 &  & $1/\sigma^2$   & 11.635 ( 0.246 ) & 12.838 ( 0.062 ) & 12.967 ( 0.069 ) & 13.006 ( 0.076 ) \\
\rowcolor{Gray} &  CUMC & c  & 0.107 ( 0.009 ) & 0.151 ( 0.009 ) & 0.151 ( 0.009 ) & 0.155 ( 0.009 ) \\
 &  & d  & 0.012 ( 0.001 ) & 0.038 ( 0.003 ) & 0.043 ( 0.003 ) & 0.044 ( 0.003 ) \\
 \rowcolor{Gray}&   & $\mu$  & 0.05 ( 0.006 ) & 0.026 ( 0.006 ) & 0.026 ( 0.006 ) & 0.025 ( 0.006 ) \\
 &  & $1/\sigma^2$   & 31.82 ( 1.495 ) & 73.354 ( 9.122 ) & 90.744 ( 7.028 ) & 93.649 ( 5.437 ) \\
\rowcolor{Gray}  & IMSG & c  & 0.161 ( 0.008 ) & 0.27 ( 0.008 ) & 0.272 ( 0.008 ) & 0.281 ( 0.008 ) \\
 &  & d  & 0.034 ( 0.002 ) & 0.1 ( 0.005 ) & 0.117 ( 0.006 ) & 0.126 ( 0.006 ) \\
 \rowcolor{Gray}&   & $\mu$  & 0.094 ( 0.005 ) & 0.042 ( 0.005 ) & 0.041 ( 0.005 ) & 0.039 ( 0.005 ) \\
 &  & $1/\sigma^2$   & 10.347 ( 0.262 ) & 17.58 ( 0.853 ) & 21.074 ( 1.368 ) & 24.011 ( 1.715 ) \\
\rowcolor{Gray}  & JMDC & c  & 0.004 ( 0.005 ) & 0.01 ( 0.005 ) & 0.012 ( 0.005 ) & 0.011 ( 0.005 ) \\
 &  & d  & 0 ( 0 ) & 0 ( 0 ) & 0 ( 0 ) & 0 ( 0 ) \\
\rowcolor{Gray} &   & $\mu$  & 0.015 ( 0.004 ) & 0.012 ( 0.004 ) & 0.011 ( 0.004 ) & 0.011 ( 0.004 ) \\
 &  & $1/\sigma^2$   & 99.602 ( 0.388 ) & 99.598 ( 0.401 ) & 99.591 ( 0.394 ) & 99.59 ( 0.4 ) \\
\rowcolor{Gray}  & MDCD & c  & 0.018 ( 0.004 ) & 0.056 ( 0.003 ) & 0.057 ( 0.003 ) & 0.066 ( 0.003 ) \\
 &  & d  & 0.032 ( 0.001 ) & 0.065 ( 0.002 ) & 0.067 ( 0.002 ) & 0.067 ( 0.002 ) \\
\rowcolor{Gray} &   & $\mu$  & 0.051 ( 0.003 ) & 0.031 ( 0.002 ) & 0.03 ( 0.002 ) & 0.027 ( 0.002 ) \\
 &  & $1/\sigma^2$   & 11.531 ( 0.13 ) & 15.159 ( 0.257 ) & 15.45 ( 0.299 ) & 15.563 ( 0.318 ) \\
\rowcolor{Gray} &  MDCR & c  & 0 ( 0.002 ) & 0.013 ( 0.002 ) & 0.013 ( 0.002 ) & 0.017 ( 0.002 ) \\
 &  & d  & 0.009 ( 0 ) & 0.021 ( 0.001 ) & 0.022 ( 0.001 ) & 0.023 ( 0.001 ) \\
\rowcolor{Gray}  &  & $\mu$  & 0.022 ( 0.001 ) & 0.014 ( 0.001 ) & 0.014 ( 0.001 ) & 0.013 ( 0.001 ) \\
 &  & $1/\sigma^2$   & 36.877 ( 0.556 ) & 52.119 ( 1.003 ) & 54.601 ( 1.179 ) & 56.045 ( 1.313 ) \\
 \rowcolor{Gray}&  NHIS NSC & c  & 0.01 ( 0.005 ) & 0.022 ( 0.005 ) & 0.022 ( 0.005 ) & 0.024 ( 0.005 ) \\
 &  & d  & 0.001 ( 0.001 ) & 0.001 ( 0.001 ) & 0.001 ( 0.001 ) & 0.001 ( 0.001 ) \\
 \rowcolor{Gray} &  & $\mu$  & 0.02 ( 0.004 ) & 0.013 ( 0.004 ) & 0.013 ( 0.004 ) & 0.011 ( 0.004 ) \\
 &  & $1/\sigma^2$   & 98.615 ( 1.334 ) & 98.633 ( 1.304 ) & 98.646 ( 1.295 ) & 98.645 ( 1.309 ) \\
\rowcolor{Gray} &  Optum & c  & -0.001 ( 0.001 ) & 0.013 ( 0.001 ) & 0.014 ( 0.001 ) & 0.019 ( 0.001 ) \\
 &  & d  & 0.014 ( 0 ) & 0.017 ( 0 ) & 0.017 ( 0 ) & 0.016 ( 0 ) \\
\rowcolor{Gray}  &  & $\mu$  & 0.03 ( 0.001 ) & 0.022 ( 0.001 ) & 0.022 ( 0.001 ) & 0.02 ( 0.001 ) \\
 &  & $1/\sigma^2$   & 16.823 ( 0.071 ) & 17.384 ( 0.102 ) & 17.313 ( 0.101 ) & 17.185 ( 0.101 ) \\
\rowcolor{Gray} &  Panther & c  & 0.155 ( 0.006 ) & 0.234 ( 0.001 ) & 0.236 ( 0.001 ) & 0.245 ( 0.001 ) \\
 &  & d  & 0.045 ( 0.01 ) & 0.054 ( 0.001 ) & 0.054 ( 0.001 ) & 0.052 ( 0.001 ) \\
\rowcolor{Gray}  &  & $\mu$  & 0.077 ( 0.004 ) & 0.042 ( 0.001 ) & 0.041 ( 0.001 ) & 0.039 ( 0.001 ) \\
 &  & $1/\sigma^2$   & 6.926 ( 0.31 ) & 7.27 ( 0.03 ) & 7.249 ( 0.03 ) & 7.206 ( 0.03 ) \\
\hline 
\end{tabular} 
 \caption{Table displaying mean and standard deviations of the prior parameters for the hypertension data set for the linear model of bias.}
    \label{tab:MeanStdDevPriors_linear_hypertension}  
\end{table*}

\begin{table*}[htp]
\centering
\scalebox{0.9}{
\begin{tabular}{lllcccc}
\hline
& & & \multicolumn{4}{c}{Effect Size} \\ 
& & & 1 & 1.5 & 2 & 4 \\ 
Training Type & Database & Prior Parameter & Mean (Std Dev) & Mean (Std Dev) & Mean (Std Dev) & Mean (Std Dev) \\ 
\hline

\rowcolor{Gray} Constant &  CCAE & Calibrated  & 2.688 ( 0.009 ) & 1.74 ( 0.004 ) & 1.72 ( 0.004 ) & 1.686 ( 0.004 ) \\
 &  & Uncalibrated  & 2.201 ( 0.01 ) & 1.054 ( 0.005 ) & 1.018 ( 0.005 ) & 0.956 ( 0.005 ) \\
\rowcolor{Gray}   & CUMC & Calibrated  & 2.957 ( 0.111 ) & 1.452 ( 0.027 ) & 1.413 ( 0.026 ) & 1.352 ( 0.025 ) \\
 &  & Uncalibrated  & 2.783 ( 0.112 ) & 1.156 ( 0.032 ) & 1.104 ( 0.031 ) & 1.025 ( 0.03 ) \\
 \rowcolor{Gray}  &IMSG & Calibrated  & 3.603 ( 0.052 ) & 2.307 ( 0.025 ) & 2.274 ( 0.025 ) & 2.225 ( 0.025 ) \\
 &  & Uncalibrated  & 3.225 ( 0.054 ) & 1.73 ( 0.031 ) & 1.68 ( 0.031 ) & 1.608 ( 0.032 ) \\
 \rowcolor{Gray}&  JMDC & Calibrated  & 1.767 ( 0.053 ) & 1.013 ( 0.031 ) & 0.981 ( 0.031 ) & 0.927 ( 0.03 ) \\
 &  & Uncalibrated  & 1.695 ( 0.055 ) & 0.91 ( 0.032 ) & 0.873 ( 0.032 ) & 0.81 ( 0.032 ) \\
\rowcolor{Gray}  & MDCD & Calibrated  & 2.876 ( 0.028 ) & 1.815 ( 0.011 ) & 1.788 ( 0.011 ) & 1.744 ( 0.011 ) \\
 &  & Uncalibrated  & 2.43 ( 0.03 ) & 1.162 ( 0.015 ) & 1.119 ( 0.014 ) & 1.043 ( 0.014 ) \\
\rowcolor{Gray} &  MDCR & Calibrated  & 2.406 ( 0.024 ) & 1.204 ( 0.007 ) & 1.174 ( 0.007 ) & 1.122 ( 0.007 ) \\
 &  & Uncalibrated  & 2.217 ( 0.025 ) & 0.915 ( 0.008 ) & 0.874 ( 0.008 ) & 0.801 ( 0.008 ) \\
\rowcolor{Gray}  & NHIS NSC & Calibrated  & 2.684 ( 0.078 ) & 1.247 ( 0.036 ) & 1.207 ( 0.036 ) & 1.131 ( 0.034 ) \\
 &  & Uncalibrated  & 2.636 ( 0.08 ) & 1.159 ( 0.037 ) & 1.114 ( 0.037 ) & 1.032 ( 0.036 ) \\
\rowcolor{Gray} &  Optum & Calibrated  & 2.35 ( 0.009 ) & 1.42 ( 0.003 ) & 1.4 ( 0.003 ) & 1.366 ( 0.003 ) \\
 &  & Uncalibrated  & 1.965 ( 0.009 ) & 0.876 ( 0.004 ) & 0.842 ( 0.004 ) & 0.781 ( 0.004 ) \\
 \rowcolor{Gray}  & Panther & Calibrated  & 3.42 ( 0.007 ) & 2.333 ( 0.004 ) & 2.304 ( 0.004 ) & 2.263 ( 0.004 ) \\
 &  & Uncalibrated  & 2.806 ( 0.007 ) & 1.435 ( 0.005 ) & 1.387 ( 0.005 ) & 1.314 ( 0.005 ) \\
\rowcolor{Gray} Linear & CCAE & Calibrated  & 2.325 ( 0.005 ) & 1.601 ( 0.003 ) & 1.632 ( 0.003 ) & 1.703 ( 0.003 ) \\
 &  & Uncalibrated  & 2.201 ( 0.01 ) & 1.054 ( 0.005 ) & 1.018 ( 0.005 ) & 0.956 ( 0.005 ) \\
 \rowcolor{Gray}  &CUMC & Calibrated  & 2.259 ( 0.034 ) & 1.118 ( 0.02 ) & 1.166 ( 0.02 ) & 1.271 ( 0.019 ) \\
 &  & Uncalibrated  & 2.783 ( 0.112 ) & 1.156 ( 0.032 ) & 1.104 ( 0.031 ) & 1.025 ( 0.03 ) \\
\rowcolor{Gray}  & IMSG & Calibrated  & 2.657 ( 0.016 ) & 1.605 ( 0.017 ) & 1.682 ( 0.016 ) & 1.846 ( 0.015 ) \\
 &  & Uncalibrated  & 3.225 ( 0.054 ) & 1.73 ( 0.031 ) & 1.68 ( 0.031 ) & 1.608 ( 0.032 ) \\
 \rowcolor{Gray} & JMDC & Calibrated  & 1.652 ( 0.044 ) & 0.987 ( 0.028 ) & 0.903 ( 0.028 ) & 0.903 ( 0.028 ) \\
 &  & Uncalibrated  & 1.695 ( 0.055 ) & 0.91 ( 0.032 ) & 0.81 ( 0.032 ) & 0.81 ( 0.032 ) \\
  \rowcolor{Gray}& MDCD & Calibrated  & 2.405 ( 0.013 ) & 1.596 ( 0.009 ) & 1.649 ( 0.009 ) & 1.757 ( 0.008 ) \\
 &  & Uncalibrated  & 2.43 ( 0.03 ) & 1.162 ( 0.015 ) & 1.119 ( 0.014 ) & 1.043 ( 0.014 ) \\
\rowcolor{Gray}  & MDCR & Calibrated  & 2.06 ( 0.012 ) & 1.119 ( 0.006 ) & 1.136 ( 0.006 ) & 1.19 ( 0.006 ) \\
 &  & Uncalibrated  & 2.217 ( 0.025 ) & 0.915 ( 0.008 ) & 0.874 ( 0.008 ) & 0.801 ( 0.008 ) \\
\rowcolor{Gray}  &NHIS NSC & Calibrated  & 2.348 ( 0.044 ) & 1.193 ( 0.032 ) & 1.158 ( 0.032 ) & 1.09 ( 0.031 ) \\
 &  & Uncalibrated  & 2.636 ( 0.08 ) & 1.159 ( 0.037 ) & 1.114 ( 0.037 ) & 1.032 ( 0.036 ) \\
 \rowcolor{Gray} & Optum & Calibrated  & 2.1 ( 0.006 ) & 1.347 ( 0.003 ) & 1.353 ( 0.003 ) & 1.371 ( 0.003 ) \\
 &  & Uncalibrated  & 1.965 ( 0.009 ) & 0.876 ( 0.004 ) & 0.842 ( 0.004 ) & 0.781 ( 0.004 ) \\
\rowcolor{Gray} &  Panther & Calibrated  & 2.564 ( 0.003 ) & 1.755 ( 0.003 ) & 1.771 ( 0.003 ) & 1.805 ( 0.003 ) \\
 &  & Uncalibrated  & 2.806 ( 0.007 ) & 1.435 ( 0.005 ) & 1.387 ( 0.005 ) & 1.314 ( 0.005 ) \\
 \rowcolor{Gray} Train on Only Negative Controls & CCAE & Calibrated  & 2.568 ( 0.01 ) & 1.583 ( 0.004 ) & 1.561 ( 0.004 ) & 1.524 ( 0.004 ) \\
 &  & Uncalibrated  & 2.201 ( 0.01 ) & 1.054 ( 0.005 ) & 1.018 ( 0.005 ) & 0.956 ( 0.005 ) \\
\rowcolor{Gray}  & CUMC & Calibrated  & 2.828 ( 0.112 ) & 1.236 ( 0.03 ) & 1.189 ( 0.029 ) & 1.117 ( 0.029 ) \\
 &  & Uncalibrated  & 2.783 ( 0.112 ) & 1.156 ( 0.032 ) & 1.104 ( 0.031 ) & 1.025 ( 0.03 ) \\
\rowcolor{Gray}  & IMSG & Calibrated  & 3.322 ( 0.054 ) & 1.892 ( 0.029 ) & 1.848 ( 0.029 ) & 1.787 ( 0.029 ) \\
 &  & Uncalibrated  & 3.225 ( 0.054 ) & 1.73 ( 0.031 ) & 1.68 ( 0.031 ) & 1.608 ( 0.032 ) \\
\rowcolor{Gray} &  JMDC & Calibrated  & 1.766 ( 0.053 ) & 1.014 ( 0.031 ) & 0.982 ( 0.03 ) & 0.928 ( 0.03 ) \\
 &  & Uncalibrated  & 1.695 ( 0.055 ) & 0.91 ( 0.032 ) & 0.873 ( 0.032 ) & 0.81 ( 0.032 ) \\
\rowcolor{Gray} &  MDCD & Calibrated  & 2.697 ( 0.029 ) & 1.569 ( 0.012 ) & 1.538 ( 0.012 ) & 1.486 ( 0.012 ) \\
 &  & Uncalibrated  & 2.43 ( 0.03 ) & 1.162 ( 0.015 ) & 1.119 ( 0.014 ) & 1.043 ( 0.014 ) \\
\rowcolor{Gray} &  MDCR & Calibrated  & 2.323 ( 0.024 ) & 1.081 ( 0.007 ) & 1.047 ( 0.007 ) & 0.989 ( 0.007 ) \\
 &  & Uncalibrated  & 2.217 ( 0.025 ) & 0.915 ( 0.008 ) & 0.874 ( 0.008 ) & 0.801 ( 0.008 ) \\
\rowcolor{Gray}  & NHIS NSC & Calibrated  & 2.689 ( 0.079 ) & 1.247 ( 0.036 ) & 1.207 ( 0.035 ) & 1.132 ( 0.034 ) \\
 &  & Uncalibrated  & 2.636 ( 0.08 ) & 1.159 ( 0.037 ) & 1.114 ( 0.037 ) & 1.032 ( 0.036 ) \\
 \rowcolor{Gray} &Optum & Calibrated  & 2.305 ( 0.009 ) & 1.36 ( 0.004 ) & 1.339 ( 0.003 ) & 1.303 ( 0.003 ) \\
 &  & Uncalibrated  & 1.965 ( 0.009 ) & 0.876 ( 0.004 ) & 0.842 ( 0.004 ) & 0.781 ( 0.004 ) \\
\rowcolor{Gray} &  Panther & Calibrated  & 3.288 ( 0.007 ) & 2.157 ( 0.004 ) & 2.126 ( 0.004 ) & 2.081 ( 0.004 ) \\
 &  & Uncalibrated  & 2.806 ( 0.007 ) & 1.435 ( 0.005 ) & 1.387 ( 0.005 ) & 1.314 ( 0.005 ) \\
\hline 
\end{tabular} 
}
 \caption{Table displaying mean and standard error of the posterior interval lengths for the hypertension data set.}
    \label{tab:MeanStdDevPriors_intervallengths_hypertension}  
\end{table*}

\begin{table*}[htp]
\centering
\begin{tabular}{lllcccc}
\hline
& & & \multicolumn{4}{c}{Effect Size} \\ 
& & & 1 & 1.5 & 2 & \multicolumn{1}{c}{4} \\ 
Training Type & Database & Calibration Type & Coverage & Coverage & Coverage &Coverage \\ 
\hline
 \rowcolor{Gray}Constant & CCAE & Calibrated  & $0.947$ & $0.943$ & $0.943$ & $0.934$ \\
 &  & Uncalibrated  & $0.818$ & $0.758$ & $0.735$ & $0.691$ \\
 \rowcolor{Gray}  &MDCD & Calibrated  & $0.959$ & $0.957$ & $0.953$ & $0.953$ \\
 &  & Uncalibrated  & $0.911$ & $0.850$ & $0.824$ & $0.792$ \\
 \rowcolor{Gray}&  MDCR & Calibrated  & $0.954$ & $0.938$ & $0.930$ & $0.924$ \\
 &  & Uncalibrated  & $0.903$ & $0.830$ & $0.804$ & $0.767$ \\
\rowcolor{Gray} & Optum & Calibrated  & $0.951$ & $0.942$ & $0.937$ & $0.929$ \\
 &  & Uncalibrated  & $0.855$ & $0.792$ & $0.761$ & $0.723$ \\
\rowcolor{Gray}Linear & CCAE & Calibrated  & $0.945$ & $0.943$ & $0.943$ & $0.942$ \\
 &  & Uncalibrated  & $0.818$ & $0.758$ & $0.735$ & $0.691$ \\
 \rowcolor{Gray} &MDCD & Calibrated  & $0.958$ & $0.956$ & $0.953$ & $0.956$ \\
 &  & Uncalibrated  & $0.911$ & $0.850$ & $0.824$ & $0.792$ \\
 \rowcolor{Gray}& MDCR & Calibrated  & $0.959$ & $0.938$ & $0.932$ & $0.940$ \\
 &  & Uncalibrated  & $0.903$ & $0.830$ & $0.804$ & $0.767$ \\
\rowcolor{Gray} & Optum & Calibrated  & $0.947$ & $0.943$ & $0.939$ & $0.939$ \\
 &  & Uncalibrated  & $0.855$ & $0.792$ & $0.761$ & $0.723$ \\
\rowcolor{Gray} Train on Only Negative Controls & CCAE & Calibrated  & $0.942$ & $0.939$ & $0.941$ & $0.927$ \\
 &  & Uncalibrated  & $0.818$ & $0.758$ & $0.735$ & $0.691$ \\
\rowcolor{Gray}  &MDCD & Calibrated  & $0.957$ & $0.956$ & $0.952$ & $0.949$ \\
 &  & Uncalibrated  & $0.911$ & $0.850$ & $0.824$ & $0.792$ \\
\rowcolor{Gray}  &MDCR & Calibrated  & $0.948$ & $0.932$ & $0.917$ & $0.898$ \\
 &  & Uncalibrated  & $0.903$ & $0.830$ & $0.804$ & $0.767$ \\
\rowcolor{Gray}  &Optum & Calibrated  & $0.944$ & $0.935$ & $0.928$ & $0.921$ \\
 &  & Uncalibrated  & $0.855$ & $0.792$ & $0.761$ & $0.723$ \\
\hline 
\end{tabular}
 \caption{Table displaying coverage for the sensitivity analysis using the depression data set.}
    \label{tab:SensitivityAnalysisDepression}  
\end{table*}

%\section{}\label{}

% \begin{figure} 
% \includegraphics{<eps-file>}% place <eps-file> in ./img  subfolder
% \caption{}
% \label{}
% \end{figure}

% \begin{table} 
% *****************
% \begin{tabular}{lll}
% \end{tabular}
% *****************
% \caption{}
% \label{}
% \end{figure}

%%%%%%%%%%%%%%%%%%%%%%%%%%%%%%%%%%%%%%%%%%%%%%
%% Supplementary Material, if any, should   %%
%% be provided in {supplement} environment  %%
%% with title and short description.        %%
%%%%%%%%%%%%%%%%%%%%%%%%%%%%%%%%%%%%%%%%%%%%%%
%\begin{supplement}
%\stitle{???}
%\sdescription{???.}
%\end{supplement}

%% ** The bibliograhy **
%\bibliographystyle{ba}
%\bibliography{Zotero.bib}

% ** Acknowledgements **
%\begin{acks}[Acknowledgments]
%\end{acks}

\end{document}